\documentclass[pdftex, pra, twocolumn, floatfix, superscriptaddress, nofootinbib]{revtex4}

\usepackage{times}
\usepackage{amsfonts}
\usepackage{amssymb}
\usepackage{amsmath}
\usepackage{graphicx}
\usepackage{subfigure}
\usepackage[colorlinks=true, breaklinks=true, linkcolor=blue, citecolor=blue, urlcolor=blue]{hyperref}
\usepackage{bbm}

\newcommand{\doi}[2]{\href{http://dx.doi.org/#1}{#2}}
\newcommand{\arxiv}[1]{\href{http://arxiv.org/abs/#1}{arXiv:#1}}
\newcommand{\urn}[1]{[\href{http://nbn-resolving.de/#1}{#1}]}

\begin{document}

\title{Momentum distributions and numerical methods for strongly interacting \\ one-dimensional spinor gases}

\author{F. Deuretzbacher}
\email{frank.deuretzbacher@itp.uni-hannover.de}
\affiliation{Institut f\"ur Theoretische Physik, Leibniz Universit\"at Hannover, Appelstrasse 2, DE-30167 Hannover, Germany}

\author{D. Becker}
\affiliation{I. Institut f\"ur Theoretische Physik, Universit\"at Hamburg, Jungiusstrasse 9, DE-20355 Hamburg, Germany}

\author{L. Santos}
\affiliation{Institut f\"ur Theoretische Physik, Leibniz Universit\"at Hannover, Appelstrasse 2, DE-30167 Hannover, Germany}

\begin{abstract}
One-dimensional spinor gases with strong $\delta$ interaction fermionize and form a spin chain. The spatial degrees of freedom of this atom chain can be described by a mapping to spinless noninteracting fermions and the spin degrees of freedom are described by a spin-chain model with nearest-neighbor interactions. Here, we compute momentum and occupation-number distributions of up to 16 strongly interacting spinor fermions and bosons as a function of their spin imbalance, the strength of an externally applied magnetic field gradient, the length of their spin, and for different excited states of the multiplet. We show that the ground-state momentum distributions resemble those of the corresponding noninteracting systems, apart from flat background distributions, which extend to high momenta. Moreover, we show that the spin order of the spin chain---in particular antiferromagnetic spin order---may be deduced from the momentum and occupation-number distributions of the system. Finally, we present efficient numerical methods for the calculation of the single-particle densities and one-body density matrix elements and of the local exchange coefficients of the spin chain for large systems containing more than 20 strongly interacting particles in arbitrary confining potentials.
\end{abstract}

\maketitle

\section{Introduction}

A one-dimensional (1D) Bose gas of spinless impenetrable point particles can be solved exactly through a simple mapping to spinless noninteracting 1D fermions \cite{Girardeau60}. Such a so-called Tonks-Girardeau gas was first realized in 2004 in experiments with ultracold atoms \cite{Kinoshita04, Paredes04}. The infinitely strong repulsion between the particles prevents the bosons from staying at the same position. As a result, the local two- and three-body correlation functions of strongly interacting bosons are substantially reduced compared to noninteracting ones \cite{LaburtheTolra04, Kinoshita05}. Moreover, the thermalization of a 1D Bose gas is substantially slower than in three dimensions \cite{Kinoshita06}.

An extremely useful feature of ultracold atoms in 1D is that the strength $g$ of the effective 1D $\delta$ interaction may be tuned to nearly arbitrary positive and negative values depending on the strength of the externally applied magnetic ($B$) field \cite{Inouye98} and of the transverse confinement \cite{Olshanii98}. More precisely, the inverse interaction strength $1/g$ may be tuned continuously from small positive values (strong repulsion) to small negative values (strong attraction) through a small change of the external $B$ field in the vicinity of a confinement-induced resonance. Thereby the ground state at $1/g>0$ evolves continuously into a highly excited metastable state at $1/g<0$ \cite{Haller09}, the so-called super Tonks-Girardeau gas \cite{Astrakharchik05}, which resembles a gas of impenetrable particles with a finite diameter.

Moreover, a few years ago, Jochim and co-workers were able to prepare a few ultracold fermionic atoms deterministically in their ground state \cite{Serwane11}. This enabled them to observe the fermionization of two distinguishable fermions in a 1D trap \cite{Zuern12}. Although the experiment could be described using the analytical solution of two $\delta$-interacting particles in a 1D harmonic trap \cite{FrankeArnold03}, there was no appropriate theoretical description for three and more particles available at that time. The reason for that was that the existing theory had previously focused on the Tonks-Girardeau ($1/g=0$) \cite{Girardeau07, Deuretzbacher08, Guan09} and/or the thermodynamic limit \cite{Guan07, Guan08, Matveev08} of multicomponent atomic gases and on the spin-incoherent Luttinger liquid regime \cite{Matveev04}. Consequently, Ref. \cite{Zuern12} stimulated an active theoretical research on this spin-1/2 few-fermion system \cite{Gharashi13, Lindgren14, Bugnion13, Sowinski13, Cui14a, Harshman14} aimed at unraveling the structure of the quasi-degenerate ground-state multiplet. Finally, a perturbative approach \cite{Volosniev14} and a spin-chain model \cite{Deuretzbacher14} have been developed for the regime around $1/g=0$. Only recently, a 1D system of fermions with large spin \cite{Pagano14}, and an antiferromagnetic Heisenberg spin chain of up to four fermions in a 1D trap \cite{Murmann15}, have been realized.

The spin-chain model has been applied to the impurity problem \cite{Levinsen15} and it has been generalized to spin-dependent interactions \cite{Volosniev15, Massignan15, Yang16a}, excited motional states \cite{Yang16b}, Bose-Fermi mixtures \cite{Hu15a}, spin-orbit coupling \cite{Cui14b, Guan15}, and $p$-wave interactions \cite{Hu15b, Yang15b}. Numerical simulations also considered a few two-component bosons \cite{Zoellner08, GarciaMarch13, GarciaMarch14a, GarciaMarch14b} and a large number of fermions \cite{Grining15} in the whole interaction regime.

Here, we present momentum and occupation-number distributions of large strongly interacting systems in different regimes. In particular, we study these distributions as a function of the spin imbalance, the strength of a $B$-field gradient, the excitations of the spin chain, the length of the particle spin, and the symmetry of the many-body wave function (fermions and bosons). We show that the antiferromagnetic ground state of strongly interacting spin-1/2 fermions can be clearly identified by means of its momentum and occupation-number distributions---in contrast to strongly interacting atoms in optical lattices. Finally, we present efficient numerical methods for the calculation of the occurring multidimensional integrals that can be applied to systems with more than 20 particles in arbitrary confining potentials.

The paper is organized as follows: Section~\ref{sec-spin-chain-model} gives an overview of the spin-chain model of strongly interacting 1D spinor gases, Sec.~\ref{sec-momentum-distributions} presents momentum and occupation-number distributions in various regimes, and Sec.~\ref{sec-numerical-methods} presents the numerical methods for the efficient calculation of the single-particle densities (Sec.~\ref{subsec-rhoi}), the one-body density matrix elements (Sec.~\ref{subsec-rhoij}), and the local exchange coefficients (Sec.~\ref{subsec-Ji}). We summarize our results in Sec.~\ref{sec-summary}.

\section{Spin-chain model for 1D strongly interacting spinor gases}
\label{sec-spin-chain-model}

\subsection{Mapping to spinless noninteracting fermions \\ and a chain of distinguishable spins}

We consider $N$ indistinguishable ultracold atoms (bosons or fermions) with spin degrees of freedom. The atoms are trapped by a spin-independent external potential~$V$ along the axial $z$ direction and they interact through a spin-independent $\delta$~potential of strength~$g$. The radial motion of the atoms is frozen to the ground state. The interaction strength~$g$ of this quasi-1D system is freely tunable through a magnetic Feshbach resonance and through the strong radial confinement \cite{Inouye98, Olshanii98}. The effective Hamiltonian of the quasi-1D system reads
\begin{equation} \label{eq-H1D}
H = \sum_i \left[ -\frac{\hbar^2}{2m} \frac{\partial^2}{\partial z_i^2} + V(z_i) \right] + g \sum_{i<j} \delta(z_i-z_j) .
\end{equation}
In the limit of infinite repulsion, ${g=\infty}$, the multicomponent system assumes properties of spinless noninteracting fermions~\cite{Girardeau60} and a chain of distinguishable noninteracting spins \cite{Deuretzbacher08}. The many-body wave functions of the ground-state multiplet may be constructed exactly through a generalization of Girardeau's Fermi-Bose mapping \cite{Girardeau60} to particles with spin and are given by \cite{Deuretzbacher08}
\begin{equation} \label{eq-map}
| \psi \rangle = \sqrt{N!} \, S_\pm \left( | \mathrm{id} \rangle | \chi \rangle \right) .
\end{equation}
$S_\pm$ is the (anti)symmetrization operator, $| \mathrm{id} \rangle$ is a spatial many-body wave function describing $N$ spinless distinguishable particles with infinite $\delta$ repulsion in a longitudinal potential $V(z)$ and ordering $z_1 < \dotsb < z_N$, and $| \chi \rangle$ is an arbitrary spin function of $N$ distinguishable spins, which describes the spin configuration of the spin chain.~\footnote{Note that besides the usual configurations, the model is also applicable to fermions with integer spin or bosons with half-integer spin.} More precisely, $S_\pm = (1/N!) \sum_P (\pm 1)^P \hat P$, where the sum runs over all permutations $P$ of $\underline N = \{1, \dotsc, N\}$, $(-1)^P$ is the sign of the permutation $P$, and the unitary operator $\hat P$ permutes particle indices, $\hat P | \alpha_1 \rangle_1 \dotsm | \alpha_N \rangle_N = | \alpha_1 \rangle_{P(1)} \dotsm | \alpha_N \rangle_{P(N)}$, where $\alpha_1, \dotsc, \alpha_N$ are quantum numbers. The spatial wave function $| \mathrm{id} \rangle$ of the spinless distinguishable particles with infinite $\delta$ repulsion is given by \cite{Deuretzbacher08} (see Appendix~\ref{app-definitions} for more details)
\begin{equation} \label{eq-id}
\langle z_1, \dotsc , z_N | \mathrm{id} \rangle = \sqrt{N!} \, \theta \left( z_1, \dotsc, z_N \right) |\psi_F| ,
\end{equation}
where $\theta (z_1, \dotsc , z_N) = 1$ if $z_1 < \dotsb < z_N$, and zero otherwise, and where $\psi_F = \det [\phi_i(z_j)]_{i,j=1,\dotsc,N} / \sqrt{N!}$ is the ground-state Slater determinant of $N$ spinless noninteracting fermions with the eigenfunctions $\phi_1(z)$, $\phi_2(z)$, $\dotsc$ of a single particle in the external potential~$V(z)$. Here, the only difference to Girardeau's mapping for spinless hard-core bosons is the additional multiplication with $\theta (z_1, \dotsc , z_N)$, which generates a wave function for distinguishable particles with particle ordering $z_1 < \dotsb < z_N$ (the factor $\sqrt{N!}$ ensures normalization). Finally, $| \chi \rangle = \sum_{m_1, \dotsc, m_N} c_{m_1, \dotsc, m_N} | m_1, \dotsc, m_N \rangle$ is an arbitrary $N$-particle spin function with $m_i$ being the spin $z$-projection quantum number of the $i$th particle. Note that although $| \mathrm{id} \rangle$ and $| \chi \rangle$ may both be nonsymmetric, application of $S_\pm$ to the product $| \mathrm{id} \rangle | \chi \rangle$ ensures that the full many-body wave function has the desired symmetry.

Equation~(\ref{eq-map}) constitutes a one-to-one correspondence between the pure spin functions~$| \chi \rangle$ of $N$ distinguishable spins and the full many-body wave functions~$| \psi \rangle$, which solve the Hamiltonian~(\ref{eq-H1D}) in the limit of infinite $\delta$~repulsion, $g=\infty$ \cite{Deuretzbacher08}. Moreover, not only spin functions~$| \chi \rangle$ may be mapped onto full many-body wave functions~$| \psi \rangle$, and vice versa, but any observable of the full continuous Hilbert space may be expressed by its counterpart in the discrete spin space. This simplifies the description of fermionized multicomponent particles substantially, as shown in the following.

\subsection{Single-particle densities}

Important experimentally measurable observables are the spin densities of the system \cite{Deuretzbacher08, Guan09, Lindgren14, Deuretzbacher14}. The density distribution of the $m$th spin component is given by \cite{Deuretzbacher08}
\begin{equation} \label{eq-rhomz}
\rho_m (z) = \sum_{i=1}^N \rho^{(i)}(z) \rho_m^{(i)}
\end{equation}
with the probability to find the $i$th particle (with whatever spin) at position~$z$,
\begin{equation} \label{eq-rhoiz}
\rho^{(i)}(z) = N! \int d z_1 \dotsi d z_N \delta (z-z_i) \theta (z_1, \dotsc , z_N) |\psi_F|^2 ,
\end{equation}
and the probability that the magnetization of the $i$th spin equals $m$,
\begin{equation} \label{eq-rhoim}
\rho_m^{(i)} = \sum_{m_1, \dotsc, m_N} \bigl| \langle m_1, \dotsc, m_N | \chi \rangle \bigr|^2 \delta_{m, m_i} .
\end{equation}
Clearly, Eq.~(\ref{eq-rhomz}) shows that the continuous spin density ${\rho_m (z)}$ is fully characterized by the $N$-tuple $\bigl( \rho_m^{(1)}, \dotsc, \rho_m^{(N)} \bigr)$.

\begin{figure}
\begin{center}
\includegraphics[width = \columnwidth]{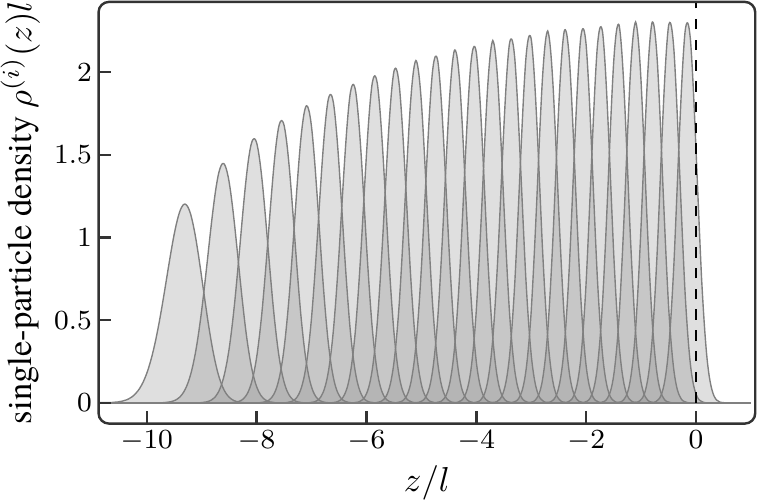}
\caption{Single-particle densities of particles 1 (left) to 25 (right) of a spin chain in a harmonic trap consisting of 50 particles. The densities 26--50 are obtained by mirroring at the vertical axis through the origin. $l$ is the harmonic-oscillator length.}
\label{fig-densities}
\end{center}
\end{figure}

The single-particle densities 1--25 of 50 harmonically trapped particles are shown in Fig.~\ref{fig-densities}. Obviously, they look like the densities of individual localized particles, as found in a Wigner crystal that is stabilized by strong longer-range interactions. The only difference is the fact that here the overlap between the densities of neighboring particles is much larger \cite{Deuretzbacher10}. One might argue that the single-particle densities~(\ref{eq-rhoiz}) are artificially constructed and not measurable and that only the spin densities~(\ref{eq-rhomz}) can be measured in the experiment. However, both densities may be identical for some spin configurations. Consider, e.g., a spin chain, in which the $i$th spin points upwards and all the other spins point downwards, $|\dotsc, \downarrow, \downarrow, \uparrow, \downarrow, \downarrow, \dotsc \rangle$. The spin density of the spin-up component then equals the $i$th particle density, thus making it visible. Similarly, one may identify individual particle densities in a spin chain in which at least two or three spin-down particles are between any pair of spin-up particles, $|\dotsc, \downarrow, \downarrow, \uparrow, \downarrow, \downarrow, \downarrow, \uparrow, \downarrow, \downarrow \dotsc \rangle$. Such spin configurations can be prepared by rotating individual spins.

Although Eq.~(\ref{eq-rhomz}) suggests that evaluating the spin densities is very simple, it actually proved to be rather difficult for large particle numbers as the calculation of the single-particle densities~(\ref{eq-rhoiz}) involves an integration over the increasingly complicated domain $z_1 < \dotsb < z_{i-1} < z < z_{i+1} < \dotsb < z_N$. Therefore, we present in Sec.~\ref{subsec-rhoi} a method that enables the calculation of the single-particle densities for large particle numbers.

\subsection{One-body density matrix elements}

The momentum and occupation-number distributions have played a central role in recent experiments \cite{Pagano14, Murmann15}. In particular, the antiferromagnetic spin state of three spin-1/2 fermions was clearly identified by means of its occupation-number distribution \cite{Murmann15}. In Sec.~\ref{sec-momentum-distributions}, we will show and discuss selected distributions in several regimes. The momentum distribution of the $m$th spin component is given by
\begin{equation} \label{eq-rhomk}
\rho_m(k) = \frac{1}{2\pi} \int dz dz' e^{\mathbbm{i} k(z-z')} \rho_m(z,z')
\end{equation}
and the mean occupancies of the $m$th spin component read
\begin{equation}
\rho_m(n) = \int dz dz' \phi_n(z) \phi_n^*(z') \rho_m(z,z') .
\end{equation}
Both distributions are calculated from the one-body density matrix of the $m$th spin component,
\begin{equation} \label{eq-rhomzz}
\rho_m(z,z') = \sum_{i,j=1}^N (\pm 1)^{i+j} \rho^{(i,j)}(z,z') \rho^{(i,j)}_m .
\end{equation}
Here, the $+$ ($-$) sign applies to bosons (fermions). Moreover, we defined the {\it spin-independent} matrix elements of the one-body density matrix,
\begin{equation} \label{eq-rhoijzz'}
\rho^{(i,j)}(z,z') = \langle\mathrm{id}| \hat\rho^{(i)}(z,z') |P_{i,\dotsc,j}\rangle
\end{equation}
with $\hat\rho^{(i)}(z,z') = |z\rangle_i \langle z'|_i$, and the matrix elements
\begin{equation} \label{eq-rhoijm}
\rho^{(i,j)}_m = \langle\chi| \hat \rho^{(i)}_m \hat P_{i,\dotsc,j} |\chi\rangle
\end{equation}
with $\hat \rho^{(i)}_m = |m\rangle_i \langle m|_i$. The loop permutation operator $\hat P_{i,\dotsc,j}$ permutes the particle indices according to the rule $i \rightarrow i+1 \rightarrow i+2 \rightarrow \dotso \rightarrow j-1 \rightarrow j \rightarrow i$ (here we assumed that $i<j$; see Appendix~\ref{app-definitions} for the full definition). The sector wave function $|P\rangle$ ($P$ is a permutation) is proportional to $|\psi_F|$ in the sector $z_{P(1)} < \dotsb < z_{P(N)}$, and zero otherwise; see Appendix~\ref{app-definitions}.

Equation~(\ref{eq-rhomzz}) is a generalization of Eq.~(\ref{eq-rhomz}) and indeed one finds $\rho_m(z) = \rho_m(z,z) = \sum_i \rho^{(i,i)}(z,z) \rho^{(i,i)}_m$.\footnote{Note that $\rho^{(i,j)}(z,z) = 0$ for $i \neq j$.} Again, the continuous spatial distribution $\rho_m(z,z')$ is fully characterized by the discrete $N^2$-tuple $\bigl( \dotsc, \rho^{(i,j)}_m, \dotsc \bigr)$. That is, once we have calculated the spin-independent one-body density matrix elements $\rho^{(i,j)}(z,z')$, we can immediately calculate the distribution $\rho_m(z,z')$ for any spin configuration $|\chi\rangle$. The calculation of the distributions $\rho^{(i,j)}(z,z')$ is, however, difficult for large particle numbers. Therefore, we present in Sec.~\ref{subsec-rhoij} a method that enables the efficient calculation of the $\rho^{(i,j)}(z,z')$ for large systems.

\begin{figure}
\begin{center}
\includegraphics[width = \columnwidth]{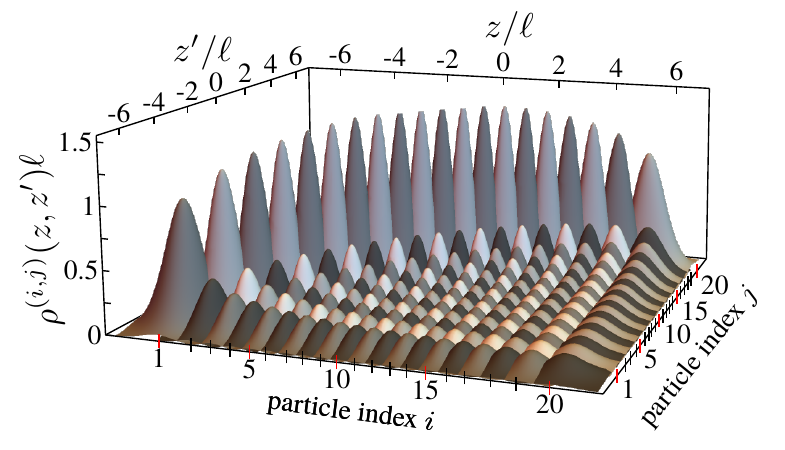}
\caption{One-body density matrix elements $\rho^{(i,j)}(z,z')$ of 20 harmonically trapped particles. $l$ is the harmonic-oscillator length.}
\label{fig-density-matrix}
\end{center}
\end{figure}

The spin-independent one-body density matrix elements $\rho^{(i,j)}(z,z')$ of 20 harmonically trapped particles are shown in Fig.~\ref{fig-density-matrix}. They resemble Gaussian-like distributions, which are located at the positions $(i,j)$ on a checkerboard. On the diagonal, $z=z'$, we recover the single-particle densities shown in Fig.~\ref{fig-densities}, which means that $\rho^{(i,i)}(z,z) = \rho^{(i)}(z)$. Also, one reads immediately from the definitions that $\rho^{(i,i)}_m = \rho^{(i)}_m$.

Equations~(\ref{eq-rhomzz})--(\ref{eq-rhoijm}) show that the shape of the one-body density matrix is directly related to the symmetry of the spin function $|\chi\rangle$ under loop permutations $P_{i,\dotsc,j}$. To become more familiar with Eqs.~(\ref{eq-rhomzz})--(\ref{eq-rhoijm}), we consider the spin-polarized case $|\chi\rangle = {|\!\uparrow, \uparrow, \uparrow, \dotsc\rangle}$. In that case, $\rho^{(i,j)}_\uparrow = 1$, $\rho_\uparrow^{(B)}(z,z') = \sum_{i,j} \rho^{(i,j)}(z,z') =: \rho^{(B)}(z,z')$ for impenetrable pointlike bosons and $\rho_\uparrow^{(F)}(z,z') = \sum_{i,j} (-1)^{i+j} \rho^{(i,j)}(z,z') =: \rho^{(F)}(z,z')$ for noninteracting fermions. That is, the one-body density matrix of spinless hard-core bosons, $\rho^{(B)}(z,z')$, is a nonalternating sum of the individual $\rho^{(i,j)}(z,z')$, while that of spinless noninteracting fermions, $\rho^{(F)}(z,z')$, is an alternating sum of all matrix elements $\rho^{(i,j)}(z,z')$. The distribution $\rho^{(B)}(z,z')$, which resembles Fig.~\ref{fig-density-matrix}, has been calculated for up to 160 particles, but the method of Ref. \cite{Papenbrock03} can unfortunately not be used for the calculation of the individual $\rho^{(i,j)}(z,z')$.

Now, consider a spin configuration $|\chi\rangle$, which is fully symmetric under any permutation of the particle indices. Then, in particular, $\hat P_{i,\dotsc,j} |\chi\rangle = |\chi\rangle$ and $\rho^{(i,j)}_m = \langle\chi| \bigl( |m\rangle_i \langle m|_i \bigr) |\chi\rangle$. Moreover, $\langle\chi| \bigl( |m\rangle_i \langle m|_i \bigr) |\chi\rangle = \langle\chi| \hat P_{i,j} |m\rangle_i \langle m|_i \hat P_{i,j} |\chi\rangle = \langle\chi| \bigl( |m\rangle_j \langle m|_j \bigr) |\chi\rangle$ and hence $\rho^{(i,j)}_m = N_m/N$, where $N_m$ is the number of particles in the $m$th spin component. Therefore, we obtain $\rho_m^{(B)}(z,z') = (N_m/N) \rho^{(B)}(z,z')$ for bosons and $\rho_m^{(F)}(z,z') = (N_m/N) \rho^{(F)}(z,z')$ for fermions. That is, the distributions have the same shape in all spin components and equal those of spinless hard-core bosons and noninteracting fermions, respectively.

Now, consider the opposite case of a fully antisymmetric spin configuration $|\chi\rangle$. Then, $\hat P_{i,\dotsc,j} |\chi\rangle = (-1)^{i+j} |\chi\rangle$, since $P_{i,\dotsc,j} = P_{i,i+1} P_{i+1,i+2} \dotsm P_{j-2,j-1} P_{j-1,j}$ is a product of $j-i$ transpositions (assuming $i<j$). Now, it follows that $\rho_m^{(B)}(z,z') = (N_m/N) \rho^{(F)}(z,z')$ for bosons and $\rho_m^{(F)}(z,z') = (N_m/N) \rho^{(B)}(z,z')$ for fermions. That is, bosons with a fully antisymmetric spin function have a one-body density matrix of spinless noninteracting fermions, and fermions with a fully antisymmetric spin function have a one-body density matrix of spinless hard-core bosons.

\subsection{Spin-chain Hamiltonian}
\label{sec-Hs}

Finally, we discuss the Hamiltonian~(\ref{eq-H1D}) in the limit of large but finite $g$. In the limit $g=\infty$, the energy eigenvalues of the system coincide with those of spinless noninteracting fermions, but the degeneracy of each level is $(2f+1)^N$ times larger ($f$ is the spin quantum number and $2f+1$ is the number of spin components), which corresponds to the number of the energetically degenerate spin configurations of the noninteracting spin chain \cite{Deuretzbacher08}. This is a direct consequence of Eq.~(\ref{eq-map}). This spin degeneracy is lifted in the limit of large but finite $g$, since nearest-neighboring spins of the spin chain now interact with each other through the spin Hamiltonian \cite{Deuretzbacher14},
\begin{equation} \label{eq-Hs}
H_s = \left( E_F - \sum_{i=1}^{N-1} J_i \right) \openone \pm \sum_{i=1}^{N-1} J_i \hat P_{i,i+1} .
\end{equation}
This Hamiltonian acts only in spin space. Its eigenfunctions $|\chi\rangle$ may be mapped onto full wave functions $|\psi\rangle$ through Eq.~(\ref{eq-map}). In Eq.~(\ref{eq-Hs}), $E_F$ is the ground-state energy of $N$ spinless noninteracting fermions,\footnote{Throughout this paper, we restrict the discussion to the ground-state multiplet. The generalization to the excited motional states is given in Ref.~\cite{Yang16b}.} $\hat P_{i,i+1}$ permutes nearest-neighboring spins, the $+$ ($-$) sign applies to fermions (bosons), and \cite{Volosniev14, Deuretzbacher14}
\begin{equation} \label{eq-Ji}
J_i = \frac{N! \hbar^4}{m^2 g} \int dz_1 \dotsi dz_N \delta (z_i-z_{i+1}) \theta (z_1, \dotsc, z_N) \left| \frac{\partial \psi_F}{\partial z_i} \right|^2
\end{equation}
are the local exchange coefficients of the interactions between nearest-neighboring spins of the spin chain. The $J_i$ are, in a good approximation, proportional to the local density cubed \cite{Deuretzbacher14, Yang16b} or, equivalently, to the local pressure within the noninteracting Fermi gas, $p(z) = \pi^2 \hbar^2 n^3(z) / (3m)$ \cite{Guan07, Guan08}. A high local pressure is accompanied by small spacings and hence a large overlap between the wave packets of neighboring particles (see Fig.~\ref{fig-densities}), which results in large local exchange coefficients. Moreover, the exchange coefficients and hence the splitting of the energy levels are linearly dependent on $1/g$. This means that the energy spectrum is inverted when a confinement-induced resonance is crossed.

The exact calculation of the $J_i$ is again difficult for large particle numbers due to the ${(N-1)}$-dimensional integrals, which have to be evaluated. Results for the $J_i$ have been presented for ${N \leq 15}$ \cite{Yang16b, Yang15a} and ${N \leq 30}$ \cite{Loft16a} particles in a harmonic trap. Recently, Loft {\it et al.} published an efficient formula and released an open source code \cite{Loft16c} for the numerical computation of the $J_i$ for $N \lesssim 35$ particles in arbitrary confining potentials \cite{Loft16b}. We show in Sec.~\ref{subsec-Ji} that this formula may be efficiently calculated using a fit with Chebyshev polynomials. The {\footnotesize MATHEMATICA} notebook containing this method and results for up to 60 harmonically trapped particles is published in the ancillary files \cite{AncillaryFiles}.

Let us consider fermions [$+$ sign in Eq.~(\ref{eq-Hs})]. The energy $E = \langle \chi | H_s | \chi \rangle$ is then minimized by a fully antisymmetric spin function, i.e., $\hat P | \chi \rangle = -| \chi \rangle$ for any permutation $P$, and assumes the value $E = E_F - 2 \sum_{i=1}^{N-1} J_i$. A fully symmetric spin function, i.e., $\hat P | \chi \rangle = | \chi \rangle$ for any permutation $P$, maximizes the energy, which assumes then the value $E = E_F$. Additionally, the symmetry of the full many-body wave function $| \psi \rangle$ requires that the ground-state spin function $| \chi \rangle$ of the fermions is combined with a fully symmetric spatial wave function, while the highest-excited state has to be combined with a fully antisymmetric one. The opposite result is obtained for bosons: A fully symmetric spin function minimizes the energy of the spin Hamiltonian $H_s$, while a fully antisymmetric one maximizes it. However, Bose symmetry requires again that the ground-state spin function has to be combined with a fully symmetric spatial wave function, while the highest-excited spin function has to be combined with a fully antisymmetric one. We conclude that in both cases, the {\it spatial} part of the many-body ground state $| \psi \rangle$ is fully symmetric under any permutation of particles.

If we have less spin components than particles available, we cannot construct a fully antisymmetric spin function, since every quantum number can at most appear once. The energy of $H_s$ is then minimized by the most antisymmetric spin function in the case of fermions and we arrive at the conclusion that now, in both cases (fermions and bosons), the {\it spatial} part of the many-body ground state is most symmetric compared to the excited states of the multiplet \cite{Fang11, Decamp16}.

\section{Momentum and occupation-number distributions}
\label{sec-momentum-distributions}

The momentum and occupation-number distributions of spinless bosons \cite{Papenbrock03}, spin-1 bosons \cite{Deuretzbacher08}, and spin-balanced \cite{Guan09, Yang15a} or nearly spin-balanced \cite{Lindgren14, Deuretzbacher14} two-component fermions have been calculated previously. Here, we will study these distributions systematically as a function of the spin imbalance, the $B$-field gradient, the excitation within the multiplet, the length of the spin, and the symmetry of the many-body wave function (fermions and bosons).

We start by discussing the limiting distributions of spinless noninteracting fermions and hard-core bosons. The momentum distribution of spinless noninteracting fermions is flat and broad, equals the density of noninteracting fermions in position space, and is given by $\rho^{(F)}(k) = \sum_{i=0}^{N-1} \phi_i^2(k)$. By contrast, the momentum distribution of spinless hard-core bosons features a narrow peak at $k=0$ \cite{Papenbrock03}. This follows immediately from Eqs.~(\ref{eq-rhomk}) and~(\ref{eq-rhomzz}), which, at $k=0$, become
\begin{equation}
\rho^{(B/F)}(k=0) = \frac{1}{2\pi} \sum_{i,j=1}^N (\pm 1)^{i+j} \int dz dz' \rho^{(i,j)}(z,z') .
\end{equation}
Clearly, since $\rho^{(i,j)}(z,z') \geq 0$ (see Fig.~\ref{fig-density-matrix}), the nonalternating sum of the bosons leads to a larger value at $k=0$ than the alternating sum of the fermions. Moreover, there is a significant probability that spinless hard-core bosons occupy high-momentum states above the Fermi edge, while these states are not occupied by spinless noninteracting fermions. This follows from the fact that the wave function of hard-core bosons exhibits symmetric cusps, $\propto |z_i-z_j|$, while the wave function of noninteracting fermions has antisymmetric zero crossings, $\propto (z_i-z_j)$, at equal particle positions, $z_i \approx z_j$ \cite{Papenbrock03}.

We conclude from these limiting cases that more symmetric spatial wave functions lead to momentum distributions with higher central peaks and more pronounced high-momentum tails, while more antisymmetric spatial wave functions lead to flatter momentum distributions and smaller high-momentum tails. The momentum and occupation-number distributions are hence a valuable measure of the symmetry of the spatial wave function. Moreover, they are also a measure of the symmetry of the spin function, since, e.g., a fully symmetric spatial wave function of a bosonic (fermionic) system has to be combined with a fully (anti)symmetric spin function \cite{Deuretzbacher08}.

\begin{figure}
\begin{center}
\includegraphics[width = 0.9 \columnwidth]{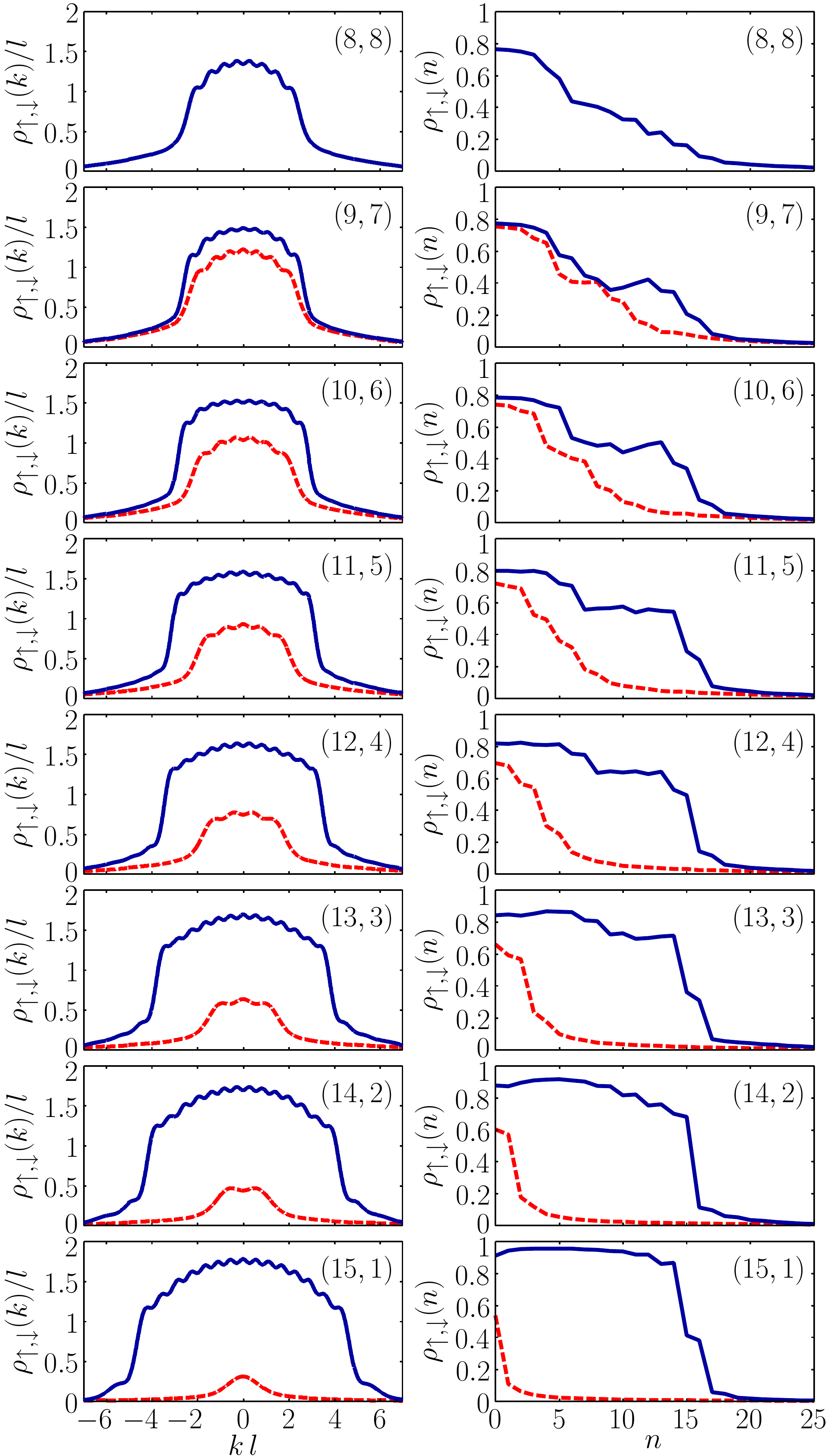}
\caption{Effect of population imbalance. Left column: Momentum distributions $\rho_\uparrow(k)$ (solid blue line) and $\rho_\downarrow(k)$ (dashed red line) of 16 spin-1/2 fermions with $(N_\uparrow, N_\downarrow) = (8,8)$ (top) to $(15,1)$ (bottom). Right column: Occupation-number distributions $\rho_\uparrow(n)$ (solid blue line) and $\rho_\downarrow(n)$ (dashed red line). $l, n$: length scale and quantum number of the harmonic oscillator.}
\label{fig-distributions-spin-imbalance}
\end{center}
\end{figure}

Let us now turn to our results. Figure~\ref{fig-distributions-spin-imbalance} shows the momentum distributions (left column) of 16 spin-1/2 fermions with infinite $\delta$ repulsion for an increasing spin imbalance (from top to bottom). One sees that the number of oscillations equals the number of particles occupying the spin component. More precisely, the central distribution resembles that of noninteracting fermions above a flat background, which extends to high momenta. For example, in the spin-balanced case, the eight noninteracting spin-up or spin-down particles occupy the eight lowest levels of the spin-up or spin-down component. Therefore, one sees the central distributions $\rho_{\uparrow, \downarrow}^{(c)}(k) \approx \sum_{i=0}^{7} \phi_i^2(k)$ above a flat background. Note, however, that the width of the central distributions $\rho_{\uparrow, \downarrow}^{(c)}(k)$ is slightly smaller than that of the distribution $\sum_{i=0}^{7} \phi_i^2(k)$, while their heights approximately coincide.

It is clear from this simple rule for the shape of the momentum distribution that the width of the central distribution of the majority component increases, while that of the minority component decreases with increasing spin imbalance. Moreover, it follows from this rule that the central distribution of the spin-balanced system has a minimum width, i.e., it cannot become as peaked as the momentum distribution of spinless hard-core bosons. Finally, we note that the population of the background of the majority particles is equally large for all spin imbalances.

The occupation-number distribution of 16 spin-1/2 fermions as a function of an increasing spin imbalance is shown in the right column of Fig.~\ref{fig-distributions-spin-imbalance}. It shows a similar dependence on the population imbalance, although there is not such a simple rule that describes its shape.

\begin{figure}
\begin{center}
\includegraphics[width = 0.9 \columnwidth]{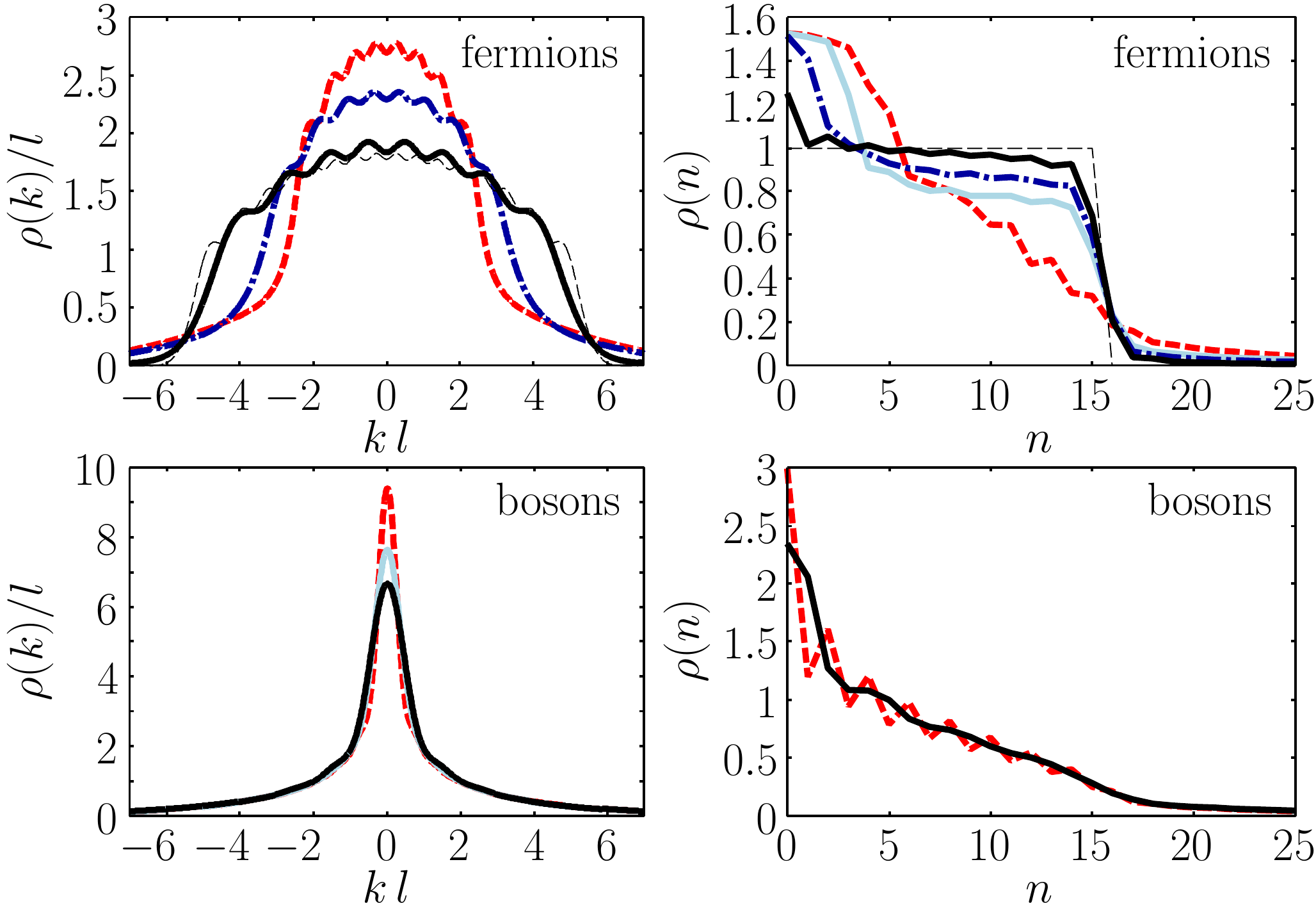}
\caption{Effect of magnetic field gradient. Top left: Momentum distributions $\rho(k) = \rho_\uparrow(k) + \rho_\downarrow(k)$ of 16 spin-balanced spin-1/2 fermions for $G/J=0$ (dashed red line), $1$ (dash-dotted blue line), and $10$ (solid black line). Top right: Occupation-number distributions $\rho(n) = \rho_\uparrow(n) + \rho_\downarrow(n)$ for $G/J=0$ (dashed red line), $1.5$ [light blue (gray) line], $3$ (dash-dotted blue line), and $10$ (solid black line). Thin dashed black lines: 16 spinless noninteracting fermions. Bottom left: Momentum distributions of 16 spin-balanced spin-1/2 bosons for $G/J=0$ (dashed red line), $0.1$ [light blue (gray) line], and $10$ (solid black line). Bottom right: Occupation-number distributions for $G/J=0$ (dashed red line) and $10$ (solid black line). $G$: strength of $B$-field gradient; $J=\sum_{i=1}^{N-1}J_i/(N-1)$: mean value of local exchange coefficients; $l, n$: length scale and quantum number of the harmonic oscillator.}
\label{fig-distributions-gradient}
\end{center}
\end{figure}

We now turn to the discussion of the effect of an increasing $B$-field gradient, which is shown in Fig.~\ref{fig-distributions-gradient}. The top left subfigure shows the momentum distribution of spin-balanced spin-1/2 fermions. One sees that the momentum distribution becomes broader and flatter with increasing $B$-field gradient, while the high-momentum tails vanish. Overall, the momentum distribution converges towards the distribution of spinless fermions in the limit of a strong $B$-field gradient, which indicates an almost antisymmetric spatial wave function.

This behavior occurs for two reasons: An increasing gradient separates the two spin components until, finally, one component is located left and one right from the trap center. The restriction to only one half of the trap volume squeezes the spin components of the position-space density and hence broadens the momentum distribution. Second, the spatial wave function within the separated spin components must be antisymmetric. Only at the boundary between the two components is the pair of unlike spins in a superposition of a singlet and a triplet spin state. Hence, only at the boundary does the spatial wave function contain a symmetric contribution. This explains the convergence of the momentum distribution towards that of spinless noninteracting fermions. The spatial wave function of a singlet of unlike spins exhibits symmetric cusps, $\propto |z_i-z_j|$, while the spatial wave function of a pair of particles with the same spin has antisymmetric zero crossings, $\propto (z_i-z_j)$, at equal particle positions, $z_i \approx z_j$ \cite{Zuern12}. The antiferromagnetic ground state of spin-1/2 fermions in the absence of a $B$-field gradient contains many pairs of unlike spins in a spin-singlet state, but there is only one pair of neighboring unlike spins at the boundary of the completely separated spin components. This explains the disappearance of the background and the high-momentum tails.

The occupation-number distribution of spin-1/2 fermions (top right in Fig.~\ref{fig-distributions-gradient}) shows the same broadening and flattening with increasing $B$-field gradient as the momentum distribution. Again, the distribution approaches the limiting distribution of spinless fermions, but it does not become completely equal to it, which is a consequence of the more symmetric spatial wave function at the boundary between the two spin components.

The momentum distribution of spin-balanced spin-1/2 hard-core bosons as a function of an increasing $B$-field gradient is shown at the bottom left in Fig.~\ref{fig-distributions-gradient}. Again, one sees a substantial flattening of the central distribution by approximately $33\%$, but its broadening is not as strong as for the fermions. By contrast, the demixing of the spin components here has approximately no effect on the occupancy of the high-momentum tails. Also, a convergence towards the momentum distribution of spinless fermions is not present.

The reason for this different behavior is that the symmetry of the spin function (and hence the spatial wave function) is changed by a strong gradient in the case of fermions, while it is not changed in the case of bosons. According to the discussion in Sec.~\ref{sec-Hs}, the ground state of Eq.~(\ref{eq-Hs}) is ferromagnetic and has a fully symmetric spin function. After the application of a $B$-field gradient, one obtains a state with separated spin components, which is still ferromagnetic and fully symmetric within each spin component. This state has hence (almost) the same permutation symmetry as before (apart from the boundary). In the initial state (zero gradient), two neighboring spins form a triplet and hence their relative spatial wave function exhibits symmetric cusps, $\propto |z_i-z_j|$. In the final state (strong gradient), a pair of like spins within the separated spin components also forms a triplet and hence the relative spatial wave function is not changed. Therefore, the symmetry of the spatial wave function is not changed by a strong $B$-field gradient. This explains the equally strong background in both cases. The minor broadening of the central peak is hence only caused by the separation of the spin components, which restricts the position-space densities of the two spin components to a smaller volume and hence broadens their momentum distributions.

\begin{figure}
\begin{center}
\includegraphics[width = \columnwidth]{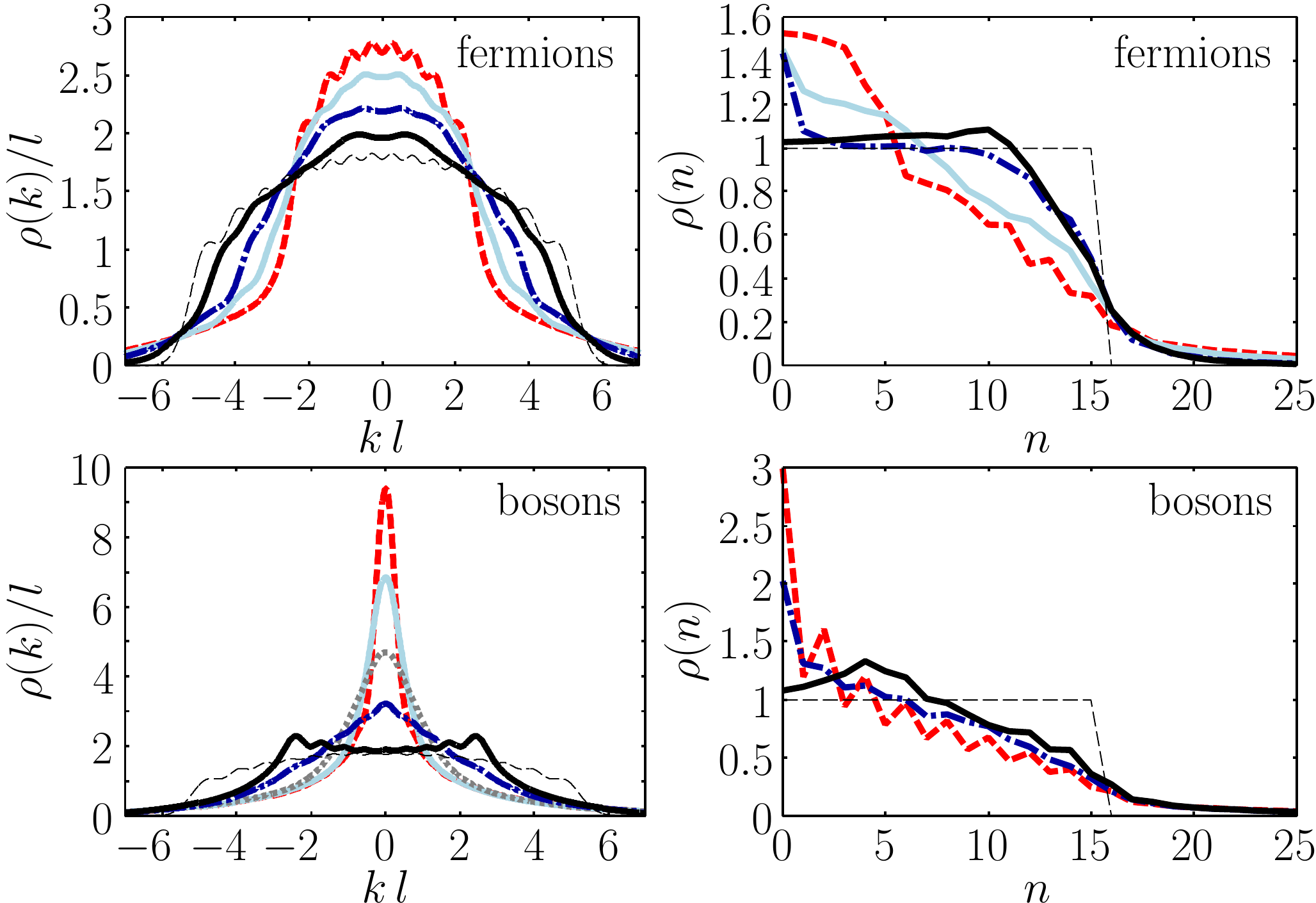}
\caption{Effect of increasing excitations. Top left: Momentum distributions $\rho(k) = \rho_\uparrow(k) + \rho_\downarrow(k)$ of 16 spin-balanced spin-1/2 fermions in the ground state (dashed red line), the 100th excited state [light blue (gray) line], the 2000th excited state (dash-dotted blue line), the 10000th excited state (solid black line), and the highest-excited state (thin dashed black line). Top right: Occupation-number distributions $\rho(n) = \rho_\uparrow(n) + \rho_\downarrow(n)$ of the ground state (dashed red line), the 1000th excited state [light blue (gray) line], the 8000th excited state (dash-dotted blue line), the 12000th excited state (solid black line), and the highest-excited state (thin dashed black line). Bottom left: Momentum distributions of 16 spin-balanced spin-1/2 bosons in the ground state (dashed red line), the 30th excited state [light blue (gray) line], the 700th excited state (dotted gray line), the 10000th excited state (dash-dotted blue line), and the highest-excited state (solid black line). Bottom right: Occupation-number distributions of the ground state (dashed red line), the 10000th excited state (dash-dotted blue line), and the highest-excited state (solid black line). Thin dashed black lines (in all subfigures): 16 spinless noninteracting fermions. $l, n$: length scale and quantum number of the harmonic oscillator.}
\label{fig-distributions-of-excited-states}
\end{center}
\end{figure}

The occupation-number distribution of spin-balanced spin-1/2 hard-core bosons for a vanishing (dashed red line) and a strong (solid black line) $B$-field gradient is shown at the bottom right in Fig.~\ref{fig-distributions-gradient}. First, we note the pronounced even-odd effect in the mean population of the harmonic-trap levels for a vanishing $B$-field gradient (dashed red line) \cite{Deuretzbacher07}. This may be viewed as a remnant of the mean-field behavior of the bosons: Weakly interacting bosons occupy together the mean-field ground state, which may be written as a superposition of harmonic-oscillator states with even parity, $\phi_\text{MF}(z) = c_0 \phi_0(z) + c_2 \phi_2(z) + c_4 \phi_4(z) + \dotsb$. The mean-field ground state of hard-core bosons is the square root of the density of spinless noninteracting fermions, $\phi_\text{MF}(z) \approx \sqrt{\rho(z)}$ \cite{Papenbrock03, Girardeau01}. This state with even parity is still much more strongly populated (by $\sqrt{N}$ bosons \cite{Papenbrock03}) than the excited natural orbitals \cite{Girardeau01}, which explains the relatively strong population of the harmonic-trap levels with even parity. This parity effect is absent when a strong $B$-field gradient is applied (solid black line), since the separated spin components are located beside the trap center, and hence the parity symmetry is broken. Apart from this parity effect, one sees again a comparatively strong population of the $n=0$ trap level for zero gradients (dashed red line). This peak is again slightly flattened and broadened when a strong $B$-field gradient is applied (solid black line).

Next, we discuss the momentum and occupation-number distributions of different states of the multiplet, shown in Fig.~\ref{fig-distributions-of-excited-states}. One sees in all cases that the ground states feature the narrowest, most peaked central distribution and the strongest population of high-momentum and high-energy states. This signals that the ground state always has the most symmetric spatial wave function among the states of the multiplet \cite{Fang11, Decamp16}, in agreement with the discussion in Sec.~\ref{sec-Hs}. As a consequence, the spin configuration of spin-1/2 fermions is most antisymmetric (antiferromagnetic) and that of spin-1/2 bosons is fully symmetric (ferromagnetic). By contrast, the highest-excited states of the multiplet always feature the broadest and flattest central distribution. This means that the highest-excited state of the multiplet always features the most antisymmetric spatial wave function. As a consequence, the spin configuration of spin-1/2 fermions is fully symmetric (ferromagnetic) and that of spin-1/2 bosons is most antisymmetric (antiferromagnetic). The other excited states interpolate continuously between these two limiting cases. Therefore, also in a large system, the antiferromagnetic states can be clearly distinguished from the ferromagnetic states by means of their momentum and occupation-number distributions.

Moreover, we note that the ground state of spin-1/2 fermions and the highest-excited state of spin-1/2 bosons feature a momentum distribution that is approximately half as broad as that of spinless fermions. The reason for that is that in both cases, the spin function cannot be fully antisymmetric, which prevents the combination with a fully symmetric (fermions) or antisymmetric (bosons) spatial wave function.

\begin{figure}
\begin{center}
\includegraphics[width = \columnwidth]{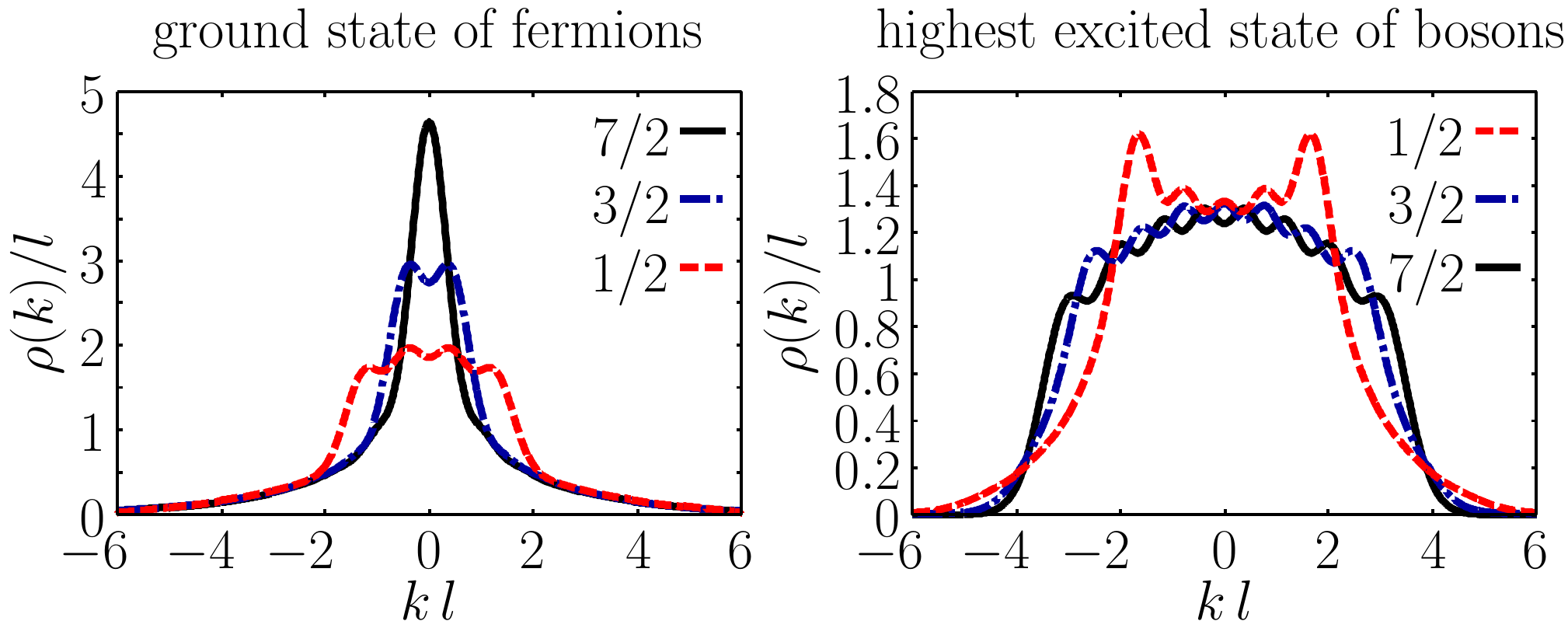}
\caption{Effect of increasing particle spin. Left: Momentum distributions $\rho(k) = \sum_m \rho_m(k)$ of eight spin-balanced fermions in the ground state and a particle spin of 1/2 (dashed red line), 3/2 (dash-dotted blue line), and 7/2 (solid black line). Right: The same for eight spin-balanced bosons in the highest-excited state. $l$ is the harmonic-oscillator length.}
\label{fig-distributions-larger-spin}
\end{center}
\end{figure}

Finally, we discuss the dependence of the momentum distribution on the particle spin, shown in Fig.~\ref{fig-distributions-larger-spin}. One sees that the momentum distribution of the fermion ground state (left) converges towards that of spinless hard-core bosons when the length of the particle spin is increased \cite{Pagano14, Yang11}. The reason for that is that spin-7/2 fermions have eight components. Hence, one can construct a fully antisymmetric spin function with eight spin-7/2 fermions. This spin function, which is the ground state of the spin Hamiltonian (\ref{eq-Hs}), can then be combined with a fully symmetric spatial wave function, which features the momentum distribution of spinless hard-core bosons.

By contrast, the momentum distribution of the highest-excited state of the bosons (right) converges towards that of spinless fermions when the length of the particle spin is increased. Here, the fully antisymmetric spin function of eight spin-7/2 bosons can be combined with a fully antisymmetric spatial wave function, which features the momentum distribution of spinless fermions.

We conclude that the momentum distributions of strongly interacting spinor fermions and bosons resemble those of their noninteracting counterparts above a flat background, which extends to high momenta. Furthermore, a change of the permutation symmetry of the ground-state spin function, induced by a $B$-field gradient or through spin excitations, leads to a dramatic change of the momentum distribution. As a result, the (anti)ferromagnetic spin order of large strongly interacting multicomponent systems can clearly be identified by means of their momentum distribution. This is impossible in a Mott insulator and a Wigner crystal, since the momentum and occupation-number distributions of strongly interacting spinless bosons and fermions are identical in these systems \cite{Deuretzbacher10}.

\section{Numerical methods}
\label{sec-numerical-methods}

We present in this section the numerical methods used to calculate the single-particle densities, one-body density matrix elements, and local exchange coefficients of large strongly interacting 1D multicomponent systems. The implementation of the formulas is given in the ancillary files~\cite{AncillaryFiles}.

\subsection{Single-particle densities}
\label{subsec-rhoi}

By evaluating the integral (\ref{eq-rhoiz}), we obtain, in a first step,
\begin{eqnarray} \label{eq-rhoiz-2}
\mspace{-20mu} \rho^{(i)}(z) & = & N! \int_{z_1 < \dotsb < z_{i-1} < z < z_{i+1} < \dotsb < z_N} \nonumber \\
& & \times d z_1 \dotsi d z_{i-1} d z_{i+1} \dotsi d z_N \nonumber \\
& & \times \bigl| \psi_F (z_1, \dotsc, z_{i-1} , z , z_{i+1}, \dotsc , z_N) \bigr|^2 .
\end{eqnarray}
This integral may be expressed by a combination of 1D integrals using $\psi_F = \det [\phi_i(z_j)]_{i,j=1,\dotsc,N} / \sqrt{N!}$ and the symmetries of $|\psi_F|^2$ (see Appendix~\ref{app-rhoi} for the derivation),
\begin{eqnarray} \label{eq-rhoiz-3}
\mspace{-30mu} \rho^{(i)}(z) & = & (-1)^{i-1} \sum_{j \leq k = 1}^N (3 \delta_{jk} - 2) \phi_j(z) \phi_k(z) \nonumber \\
& & \times \sum_{l=i-1}^{N-1} (-1)^l \binom{l}{i-1} \sum_{p \in P(j,k,l)} \det A_p(z) .
\end{eqnarray}
Here, $P(j,j,l)$ is the set of all $l$-dimensional ordered subsets of $\underline N \setminus \{j\}$ and $P(j,k,l)$ is the set of all $l$-dimensional ordered subsets of $\underline N \setminus \{j\}$, which contain $k$. The matrix $A_p$ is defined by $A_p = (A_{ij})_{i \in p, j \in p}$ for $p \in P(j,j,l)$ and $A_p = (A_{ij})_{i \in p, j \in \overline p}$ for $p \in P(j,k,l)$, where $\overline p$ is obtained from $p$ by replacing $k$ by $j$. The matrix elements $A_{ij}$ are given by the 1D integrals $A_{ij}(z) = \int_{-\infty}^z dx \, \phi_i(x) \phi_j(x)$ with the eigenfunctions $\phi_1$, $\phi_2$, $\dotsc$ of the external potential~$V$. Additionally, we defined $\sum_{p \in P(j,k,0)} \det A_p = \delta_{jk}$.

At first glance, it may seem that the computational costs of evaluating Eq.~(\ref{eq-rhoiz-3}) scale exponentially with $N$ due to the sums
\begin{equation} \label{eq-def-Aklm}
\mathcal{A}_{k \ell m} (z) := (-1)^m \sum_{p \in P(k,\ell,m)} \det A_p(z)
\end{equation}
over permutations $p$, where $\mathcal{A}_{k \ell 0} = \delta_{k \ell}$ and $\mathcal{A}_{k \ell,N-1} = 0$ for $k \ne \ell$. No matter the choice of $N$ and $i$, however, individual terms $\det A_p(z)$ never appear separately in Eq.~(\ref{eq-rhoiz-3}). Instead, it is sufficient to know the sums $\mathcal{A}_{k \ell m} (z)$ as a whole to obtain any of the $\rho^{(i)} (z)$. The former can, in turn, be easily and efficiently calculated numerically using the relation
\begin{eqnarray} \label{eq-gen-func-1}
G_{k\ell}^z(x) & := & (-1)^{k+\ell+\delta_{k\ell} + N} \det \langle (x A_{ij})(z) - \openone_N \rangle_{k\ell} \nonumber \\
& = & \sum_{m = 0}^{N-1} x^m \mathcal{A}_{k \ell m}(z) ,
\end{eqnarray}
where $\openone_N$ is the $N$-dimensional identity matrix and $\langle \cdot \rangle_{k \ell}$ denotes the matrix operation of deleting row $k$ and column $\ell$. Hence, Eq.~\eqref{eq-gen-func-1} defines a generating function for exactly those determinant sums $\mathcal{A}_{k\ell m}(z) = (m!)^{-1} d^m G_{k\ell}^z(x)/d x^m \vert_{x=0}$ needed to obtain $\rho^{(i)}(z)$. With $G_{k\ell}^z(x)$, we get access to these sums essentially by calculating the determinant of a single $(N-1)$-dimensional matrix per $x$.

It is easy to see that the absolute values of two terms $\mathcal{A}_{k \ell m}(z)$ and $\mathcal{A}_{k \ell m'}(z)$ can differ by many orders of magnitude if $|m - m'| \gg 1$. Therefore, straightforward numerical approaches, such as, e.g., using finite differences to evaluate the higher-order derivatives of $G_{k\ell}^z(x)$ at $x = 0$, are likely to become numerically unstable for $N$ larger than about $10$.

Instead, we use Chebyshev polynomials of the first kind to numerically obtain an expression of the right-hand side of Eq.~(\ref{eq-gen-func-1}) by fitting a polynomial to the generating function in the range $x \in [-R,R]$, where parameter $R \ge 1$ allows adjusting where in the range $1 \le m \le N$ the fit yields highest accuracy. The fit method itself is a well-known, numerically stable, and efficient procedure \cite{Press07}. We just state the result here, which is given by
\begin{equation} \label{eq-approx-gen-func}
G_{k \ell}^z(x) \approx \delta_{k, \ell} + \frac{2}{M} \sum_{p,q \in \underline M} \gamma_{p,q} G_{k \ell}^z(\Gamma_{1,q}) T_{p} (x/R)
\end{equation}
with $\gamma_{p,q} := \cos [ \pi p (2q - 1)/(2M)]$ and $\Gamma_{p,q} := R \gamma_{p,q}$, where $M \ge N$ and $T_p(x) = \cos[ p  \arccos ( x )]$ is the $p$th Chebyshev polynomial of the first kind. Note that while evaluating Eq.~(\ref{eq-approx-gen-func}) requires only $M$ evaluations of $G$ (one for each $\Gamma_{1,q}$), the resulting fit is valid (within some accuracy bound) for all $x \in [-R,R]$. 

With the fit polynomial~(\ref{eq-approx-gen-func}) given, the $\mathcal{A}_{k \ell m} (z)$ are then approximated by
\begin{equation} \label{eq-fit-Aklm}
\mathcal{A}_{k \ell m} (z) \approx \frac{2}{M R^m} \sum_{p,q \in \underline M} c_{p,m} \gamma_{p,q} G_{k \ell}^z(\Gamma_{1,q})
\end{equation}
for $l > 0$. Here, $c_{p,m}$ denotes the $m$th-order coefficient of $T_p (x)$. This approximate relation becomes (analytically) exact for $M, R \to \infty$. For $N \lesssim 30$, however, it already yields results with a relative precision of about $10^{-4}$ for $M = N$ and $R = 2$. Results of any desired (higher) accuracy (at higher computational costs) can be obtained by performing separate fits for multiple values of $R$, while checking for convergence of the right-hand side of Eq.~(\ref{eq-fit-Aklm}) as a function of $R$ and, to a lesser extend, of $M$.

Depending on the number of particles and desired fit accuracy, however, it might be additionally required to employ a floating-point arithmetic that exceeds the native machine precision ($\approx$16 decimal digits for 64-bit floating point numbers). For example, to obtain the density of 50 particles with an (absolute) accuracy of $10^{-4}$, we had to increase the floating-point precision to 60 decimal digits~\cite{AncillaryFiles}.

\subsection{One-body density matrix elements}
\label{subsec-rhoij}

The evaluation of the matrix element (\ref{eq-rhoijzz'}) yields for $i<j$ the $(N-1)$-dimensional integral
\begin{widetext}
\begin{eqnarray}
\rho^{(i,j)}(z,z') & = & N! \, \theta(z,z') \int_{z_1 < \dotsb < z_{i-1} < z < z_{i+1} < \dotsb < z_j < z' < z_{j+1} < \dotsb < z_N} dz_1 \dotsi dz_{i-1} dz_{i+1} \dotsi d z_N \nonumber \\
& & \times \bigl| \psi_F (z_1, \dotsc, z_{i-1}, z, z_{i+1}, \dotsc, z_N) \psi_F (z_1, \dotsc, z_{i-1}, z', z_{i+1}, \dotsc, z_N) \bigr| .
\end{eqnarray}
Performing a similar calculation as in Appendix~\ref{app-rhoi}, one finds, for $i \leq j$ (see the ancillary files~\cite{AncillaryFiles} for the derivation),
\begin{equation} \label{eq-rhoijzz'-2}
\rho^{(i,j)}(z,z') = \theta(z,z') \!\! \sum_{k,l = 1}^N \sum_{m=i-1}^{j-1} \sum_{n = j-m-1}^{N-i} \!\! \left( 2 \delta_{k,l} - 1 \right) \! \phi_k(z) \phi_l(z') (-1)^n \binom{m}{m+1-i} \! \binom{n}{j-m-1} \!\! \sum_{p \in P(k,l,m,n)} \!\! \det A_p (z,z') .
\end{equation}
\end{widetext}
Here, $P(k,k,m,n)$ is the set of pairs of ordered $m$- and $n$-tuples $p=(i_1, \dotsc, i_m)(j_1, \dotsc, j_n)$ with $\{ i_1, \dotsc, i_m \}$, $\{ j_1, \dotsc, j_n \} \subset \underline N \setminus \{ k \}$, and $\{ i_1, \dotsc, i_m \} \cap \{ j_1, \dotsc, j_n \} = \emptyset$. $p$ are the row and column indices of $A_p$. $(i_1, \dotsc, i_m)$ and $(j_1, \dotsc, j_n)$ are the row indices of matrix elements $A_{ij}(z)$ and $A_{ij}(z')$, respectively. $P(k,l,m,n)$ is the set of pairs of ordered $m$- and $n$-tuples $p=(i_1, \dotsc, i_m)(j_1, \dotsc, j_n)$ with disjoint sets $\{ i_1, \dotsc, i_m \}$ and $\{ j_1, \dotsc, j_n \}$, which contain elements of $\underline N \setminus \{ k, l \}$ and where one of the sets also contains the element $l$. The column indices $\overline p$ are obtained from $p$ by replacing $l$ by $k$. Additionally, we define $\sum_{p \in P(k,l,0,0)} \det A_p = \delta_{kl}$. Using the symmetry $\rho^{(i,j)}(z,z') = \rho^{(j,i)}(z',z)$, one may also calculate the matrix elements with the indices $i \geq j$ in the domain $z > z'$.

Similar to Sec.~\ref{subsec-rhoi}, it is possible to define a generating function $G_{k \ell}^{zz'} (x,y) := \sum_{m,n = 0}^{N-1} \mathcal{A}_{k\ell mn} (z,z') x^m y^n$ for the determinant sums,
\begin{equation}
\mathcal{A}_{k \ell m n} (z,z') := (-1)^{m+n} \sum_{p \in P(k,\ell,m,n)} \det A_p (z, z') ,
\end{equation}
that appear in Eq.~(\ref{eq-rhoijzz'-2}) and are essential to calculate $\rho^{(i,j)}(z,z')$. Just as the density matrix, this generating function depends on two real parameters ($x$ and $y$). It is given by
\begin{equation}
G_{k \ell}^{z z'}(x,y) = \frac{1}{1+\delta_{k\ell}} \sum_{\alpha = 1,2} \det (\mathbf{A}(x,y) - \mathbf{I}_\alpha)
\end{equation}
with block matrices 
\begin{equation}
\mathbf{A}(x,y) :=
\begin{pmatrix}
\langle (x A_{ij})(z) \rangle_{k\ell} & \langle (x A_{ij})(z) \rangle_{k\ell} \\
\langle (y A_{ij})(z') \rangle_{k\ell} & \langle(y A_{ij})(z') \rangle_{k\ell}
\end{pmatrix}
\end{equation}
and
\begin{equation}
\mathbf{I}_\alpha :=
\begin{pmatrix}
\langle \openone_N + \delta_{\alpha,1} \Delta_{\ell,k} \rangle_{k\ell} & 0 \\
0 & \langle \openone_N + \delta_{\alpha,2} \Delta_{\ell,k} \rangle_{k\ell}
\end{pmatrix} ,
\end{equation}
where matrix $\Delta_{\ell,k} = ( \delta_{i\ell} \delta_{jk} )_{i,j \in \underline N}$ has only one nonzero element in row $\ell$ and column $k$. Again, this generating function can be effectively evaluated using a (two-dimensional) fit based on Chebyshev polynomials~\cite{AncillaryFiles}. This yields
\begin{eqnarray}
& & \mspace{-55mu} \mathcal{A}_{k\ell m n} (z,z') \nonumber \\
& & \mspace{-55mu} \approx \frac{4}{M^2 R^{m+n}} \sum_{ \substack{\lambda,\mu\\ \alpha,\beta} \in \underline M} c_{\lambda,m} c_{\mu,n} \gamma_{\lambda,\alpha} \gamma_{\mu,\beta} G_{jk}^{zz'}(\Gamma_{1,\alpha}, \Gamma_{1,\beta})
\end{eqnarray}
for $m, n > 0$. Furthermore, $\mathcal{A}_{k\ell m 0} (z,z') = \mathcal{A}_{k\ell m} (z)$ and $\mathcal{A}_{k\ell 0 n} (z,z') = \mathcal{A}_{k\ell n} (z')$ as given by Eq.~(\ref{eq-fit-Aklm}).

\subsection{Exchange coefficients}
\label{subsec-Ji}

The exchange coefficients may be efficiently calculated using the formula~\cite{Loft16c}
\begin{eqnarray} \label{eq-Ji-2}
J_i & \!\! = & \!\! (-1)^{N-i} \frac{\hbar^4}{m^2 g} \sum_{j \leq k=1}^N \sum_{l = 0}^{N-1-i} (2 - \delta_{jk}) \frac{(-1)^{j+k}}{l!} \nonumber \\
& & \!\! \times \binom{N-l-2}{i-1} \int_{-\infty}^{+\infty} dz \, \Bigl( \phi_j''(z) \phi_k'(z) + \phi_j'(z) \phi_k''(z) \Bigr) \nonumber \\
& & \!\! \times \left[ \frac{\partial^l}{\partial\lambda^l} \det \, \Bigl\langle A(z) - \lambda \openone_N \Bigr\rangle_{jk} \right]_{\lambda = 0}
\end{eqnarray}
with the $N \times N$ matrix $A(z) = [A_{ij}(z)]_{i,j=1,\dotsc,N}$, the $N \times N$ identity matrix $\openone_N$, and $\langle \cdot \rangle_{jk}$ denoting the matrix operation of deleting the $j$th row and the $k$th column. We derive Eq.~(\ref{eq-Ji-2}) in Appendix~\ref{app-Ji} using a similar formula for the $\rho^{(i)}(z)$~\cite{Deuretzbacher08, Deuretzbacher09}.

The $d$th partial derivative terms of $(A(z) - x \openone_N)$ can be identified, up to a sign, with the determinant sums $\mathcal{A}_{k \ell, N-m-1} (z)$ defined in \eqref{eq-def-Aklm} using the relation
\begin{equation}\label{eq-switch-d}
\frac{\partial^k}{\partial x^k} (\mathcal{M} - x \openone_M) \bigr\vert_{x = 0} = \frac{k!}{(M-k)!}\frac{\partial^{M-k}}{\partial x^{M-k}} (x \mathcal{M} - \openone_M) \bigr\vert_{x = 0},
\end{equation}
where $\mathcal{M}$ is an arbitrary matrix of dimension $M$. Hence, by plugging \eqref{eq-switch-d} into \eqref{eq-Ji-2}, we arrive at
\begin{equation}
\begin{split}
J_i & = \frac{(-1)^i \hbar^4}{m^2 g} \hspace*{-1mm} \sum_{k \leq \ell = 1}^N \sum_{d=i}^{N-1} (2 - 3\delta_{k\ell}) \binom{d-1}{i-1} \\
 & \phantom{=\;} \times \int_{-\infty}^{\infty} \hspace*{-3mm} dz \Bigl( \phi_k''(z) \phi_\ell'(z) + \phi_k'(z) \phi_\ell''(z) \Bigr) \mathcal{A}_{k \ell d} (z) .
\end{split}
\end{equation}
Just as with the particle density, we can employ approximation \eqref{eq-fit-Aklm} to evaluate this equation efficiently~\cite{AncillaryFiles}.

\section{Summary}
\label{sec-summary}

We calculated momentum and occupation-number distributions of large systems of strongly interacting 1D spinor gases in different regimes. We found that the momentum distributions of strongly interacting spinor fermions and bosons resemble those of their noninteracting counterparts above a flat background. Furthermore, we found that the momentum distributions change dramatically when the permutation symmetry of the ground-state spin function is changed, e.g., by a $B$-field gradient or by exciting the system. As a result, (anti)ferromagnetic spin order of large strongly interacting spinor gases can clearly be identified by means of their momentum distributions. This should be contrasted with Mott insulators or Wigner crystals, where the spin order has no impact on the momentum distribution. Furthermore, we presented efficient methods for the numerical calculation of the spin-independent single-particle densities and one-body density matrix elements and the local exchange coefficients of large systems of strongly interacting 1D spinor gases.

\vspace{1mm}

\section*{\uppercase{Acknowledgments}}

This work was supported by the DFG (Projects No. SA 1031/7-1 and No. RTG 1729) and the Cluster of Excellence QUEST.

\begin{appendix}

\section{Definitions}
\label{app-definitions}

The action of a permutation operator $\hat P$ on a many-body state $|\alpha_1, \dotsc, \alpha_N\rangle$ is defined by
\begin{eqnarray} \label{eq-P-acts-on-state}
\hat P |\alpha_1, \dotsc, \alpha_N\rangle & = & \hat P |\alpha_1\rangle_1 \dotsm |\alpha_N\rangle_N \nonumber \\
& = & |\alpha_1\rangle_{P(1)} \dotsm |\alpha_N\rangle_{P(N)} .
\end{eqnarray}
The action of $\hat P$ on a spin function is hence given by
\begin{eqnarray}
\hat P |m_1, \dotsc, m_N\rangle & = & |m_1\rangle_{P(1)} \dotsm |m_N\rangle_{P(N)} \nonumber \\
& = & |m_{P^{-1}(1)}\rangle_1 \dotsm |m_{P^{-1}(N)}\rangle_N \nonumber \\
& = & |m_{P^{-1}(1)}, \dotsc, m_{P^{-1}(N)}\rangle .
\end{eqnarray}
We use the cycle notation to specify a permutation. For example, the permutation $P_{\alpha,\beta,\gamma}$ permutes the particle indices according to the prescription $\alpha \rightarrow \beta \rightarrow \gamma \rightarrow \alpha$. The identity permutation is denoted by ``$\mathrm{id}$.'' The loop permutation is defined by
\begin{equation}
P_{i,\dotsc,j} = \left\{
\begin{aligned}
& P_{i,i+1,\dotsc,j-1,j} \quad \text{for} \quad i<j \\
& \mathrm{id} \mspace{106.5mu} \text{for} \quad i=j \\
& P_{i,i-1,\dotsc,j+1,j} \quad \text{for} \quad i>j .
\end{aligned}
\right.
\end{equation}
We define nonsymmetric spatial sector wave functions,
\begin{equation} \label{eq-sector-wf}
\langle z_1, \dotsc , z_N | P \rangle = \sqrt{N!} \, \theta \left( z_{P(1)}, \dotsc, z_{P(N)} \right) |\psi_F| ,
\end{equation}
where $\theta \left( z_{P(1)}, \dotsc, z_{P(N)} \right) = 1$ if $z_{P(1)} < \dotsb < z_{P(N)}$, and zero otherwise, and where $\psi_F = \det [\phi_i(z_j)]_{i,j=1,\dotsc,N} / \sqrt{N!}$ is the ground-state Slater determinant of $N$ spinless noninteracting fermions with the eigenfunctions $\phi_1(z)$, $\phi_2(z)$, $\dotsc$ of a single particle in the external potential~$V(z)$. The sector wave functions $|P\rangle$ are therefore proportional to $|\psi_F|$ in the sector $z_{P(1)} < \dotsb < z_{P(N)}$, and zero otherwise. The sector wave function $|\mathrm{id}\rangle$, defined in Eq.~(\ref{eq-id}), is the special case belonging to the identity permutation. The sector wave functions are orthonormal, i.e., $\langle P | P' \rangle = \delta_{P,P'}$. The action of a permutation operator $\hat P$ on a sector wave function $|P'\rangle$ is given by
\begin{equation} \label{eq-P-acts-on-sector-wf}
\hat P |P'\rangle = |P \circ P'\rangle .
\end{equation}
This follows from
\begin{widetext}
\begin{subequations}
\begin{align}
& \hat P \int dz_1 \dotsi dz_N | z_1, \dotsc, z_N \rangle \langle z_1, \dotsc, z_N | P' \rangle \nonumber \\
& \mspace{10mu} = \int dz_1 \dotsi dz_N | z_1 \rangle_{P(1)} \dotsm | z_N \rangle_{P(N)} \sqrt{N!} \, \theta \left( z_{P'(1)}, \dotsc, z_{P'(N)} \right) |\psi_F| \label{eq-PP'1} \\
& \mspace{10mu} = \int dz_{P(1)} \dotsi dz_{P(N)} | z_{P(1)} \rangle_{P(1)} \dotsm | z_{P(N)} \rangle_{P(N)} \sqrt{N!} \, \theta \left( z_{P \circ P'(1)}, \dotsc, z_{P \circ P'(N)} \right) |\psi_F| \label{eq-PP'2} \\
& \mspace{10mu} = \int dz_1 \dotsi dz_N | z_1, \dotsc, z_N \rangle \sqrt{N!} \, \theta \left( z_{P \circ P'(1)}, \dotsc, z_{P \circ P'(N)} \right) |\psi_F| \label{eq-PP'3} \\
& \mspace{10mu} = \int dz_1 \dotsi dz_N | z_1, \dotsc, z_N \rangle \langle z_1, \dotsc, z_N | P \circ P' \rangle . \label{eq-PP'4}
\end{align}
\end{subequations}
The first step, given by Eq.~(\ref{eq-PP'1}), follows from the definitions~(\ref{eq-P-acts-on-state}) and~(\ref{eq-sector-wf}); the second step, given by Eq.~(\ref{eq-PP'2}), follows from the renaming $z_1 \rightarrow z_{P(1)}$, $\dotsc$, $z_N \rightarrow z_{P(N)}$ and the fact that $|\psi_F|$ is symmetric under any permutation of its arguments; the third step, given by Eq.~(\ref{eq-PP'3}), follows from a change of the order of integration and of the kets in the tensor product $| z_{P(1)} \rangle_{P(1)} \dotsm | z_{P(N)} \rangle_{P(N)}$; and the last step follows again from the definition (\ref{eq-sector-wf}).

The operator that measures the spin-independent density of the $i$th particle is defined by
\begin{equation}
\hat \rho^{(i)}(z) = |z\rangle_i \langle z|_i = \int d z_1 \dotsi d z_N \delta(z-z_i) | z_1, \dotsc, z_N \rangle \langle z_1, \dotsc, z_N | \, .
\end{equation}
The operator for the probability that the $i$th spin has magnetization $m$ is defined by
\begin{equation}
\hat \rho^{(i)}_m = |m\rangle_i \langle m|_i = \sum_{m_1, \dotsc, m_N} \delta_{m,m_i} |m_1, \dotsc, m_N\rangle \langle m_1, \dotsc, m_N | \, . \mspace{40mu}
\end{equation}
The operator of the spin-independent one-body density matrix of the $i$th particle is defined by
\begin{equation}
\hat \rho^{(i)}(z,z') = |z\rangle_i \langle z'|_i = \int d z_1 \dotsi dz_{i-1} dz_{i+1} \dotsi d z_N | z_1, \dotsc, z_{i-1}, z, z_{i+1}, \dotsc, z_N \rangle \langle z_1, \dotsc, z_{i-1}, z', z_{i+1}, \dotsc, z_N | \, .
\end{equation}
\end{widetext}

\section{Single-particle densities}
\label{app-rhoi}

Here, we derive Eq.~(\ref{eq-rhoiz-3}) from Eq.~(\ref{eq-rhoiz-2}). The calculation resembles that of Ref. \cite{Deuretzbacher09}. But first, we derive Eqs.~(\ref{eq-rhomz})--(\ref{eq-rhoim}). The observable for measuring a particle at position $z$ in the $m$th spin component is given by
\begin{equation}
\hat \rho_m(z) = \sum_{i=1}^N |z,m\rangle_i \langle z,m|_i = \sum_{i=1}^N \hat \rho^{(i)}(z) \hat \rho^{(i)}_m
\end{equation}
with $\hat \rho^{(i)}(z) = |z\rangle_i \langle z|_i$ the density of the $i$th particle and $\hat \rho^{(i)}_m = |m\rangle_i \langle m|_i$ the probability that the $i$th spin has magnetization $m$. The expectation value of $\hat \rho_m(z)$ for a system being in state $| \psi \rangle = \sqrt{N!} S_\pm | \mathrm{id} \rangle | \chi \rangle$ is given by
\begin{subequations}
\begin{align}
\langle \hat \rho_m(z) \rangle & = N! \langle \chi | \langle \mathrm{id} | \hat \rho_m(z) S_\pm | \mathrm{id} \rangle | \chi \rangle \label{eq-rhomz-2a} \\
& = \sum_i \sum_P (\pm 1)^P \langle\mathrm{id}| \hat \rho^{(i)}(z) |P\rangle \langle\chi| \hat \rho^{(i)}_m \left( \hat P|\chi\rangle \right) \label{eq-rhomz-2b} \\
& = \sum_i \langle\mathrm{id}| \hat \rho^{(i)}(z) |\mathrm{id}\rangle \langle\chi| \hat \rho^{(i)}_m |\chi\rangle . \label{eq-rhomz-2c}
\end{align}
\end{subequations}
Here, we used $S_\pm^\dagger = S_\pm$, $S_\pm \hat \rho_m(z) = \hat \rho_m(z) S_\pm$, and $S_\pm^2 = S_\pm$ in the first step, given by Eq.~(\ref{eq-rhomz-2a}); $S_\pm = (1/N!) \sum_P (\pm 1)^P \hat P$ and $\hat P|\mathrm{id}\rangle = |P\rangle$ [see Eq.~(\ref{eq-P-acts-on-sector-wf})] in the second step, given by Eq.~(\ref{eq-rhomz-2b}); and the fact that different sector wave functions have no overlap, $\langle\mathrm{id}| \hat \rho^{(i)}(z) |P\rangle = \delta_{\mathrm{id},P} \langle\mathrm{id}| \hat \rho^{(i)}(z) |\mathrm{id}\rangle$, in the last step, given by Eq.~(\ref{eq-rhomz-2c}). Using $\langle\mathrm{id}| \hat \rho^{(i)}(z) |\mathrm{id}\rangle = \rho^{(i)}(z)$ and $\langle\chi| \hat \rho^{(i)}_m |\chi\rangle = \rho^{(i)}_m$, we obtain Eqs.~(\ref{eq-rhomz})--(\ref{eq-rhoim}).

Next, we decompose the $(N-1)$-dimensional integral~(\ref{eq-rhoiz-2}) into $(N-1)$ 1D integrals. First, we extend the domain of integration from $z_1 < \dotsb < z_{i-1} < z < z_{i+1} < \dotsb < z_N$ to $z_1, \dotsc, z_{i-1} < z < z_{i+1}, \dotsc, z_N$. We can do this since the integrand $|\psi_F (z_1, \dotsc, z_{i-1}, z, z_{i+1}, \dotsc, z_N)|^2$ is symmetric under any permutation of the first $i-1$ variables $z_1, \dotsc, z_{i-1}$ and the last $N-i$ variables $z_{i+1}, \dotsc, z_N$. We have to divide by the factor $(i-1)!(N-i)!$ since the last volume is by this factor larger than the first volume. We obtain, from Eq.~(\ref{eq-rhoiz-2}),
\begin{widetext}
\begin{equation}
\rho^{(i)}(z) = \frac{N!}{(i-1)!(N-i)!} \int_{-\infty}^z d z_1 \dotsi \int_{-\infty}^z d z_{i-1} \int_z^\infty d z_{i+1} \dotsi \int_z^\infty d z_N |\psi_F (z_1, \dotsc, z_{i-1}, z, z_{i+1}, \dotsc, z_N)|^2 .
\end{equation}
Inserting the Leibniz formula for the Slater determinant,
\begin{equation}
\psi_F = \frac{1}{\sqrt{N!}} \sum_{P \in S_N} (-1)^P \prod_{i=1}^N \phi_{P(i)}(z_i) ,
\end{equation}
we obtain
\begin{eqnarray} \label{eq-rhoiz-4}
\rho^{(i)}(z) & = & \frac{1}{(i-1)!(N-i)!} \int_{-\infty}^z d z_1 \dotsi \int_{-\infty}^z d z_{i-1} \int_z^\infty d z_{i+1} \dotsi \int_z^\infty d z_N \sum_{P \in S_N} \sum_{P' \in S_N} (-1)^P (-1)^{P'} \nonumber \\
& & \times \phi_{P(1)}(z_1) \phi_{P'(1)}(z_1) \dotsm \phi_{P(i)}(z) \phi_{P'(i)}(z) \dotsm \phi_{P(N)}(z_N) \phi_{P'(N)}(z_N)
\end{eqnarray}
and using the definitions
\begin{eqnarray}
A_{ij}(z) & = & \int_{-\infty}^z dx \, \phi_i(x) \phi_j(x) , \\
B_{ij}(z) & = & \int_z^\infty dx \, \phi_i(x) \phi_j(x) = \delta_{ij} - A_{ij}(z)
\end{eqnarray}
we get
\begin{equation}
\rho^{(i)} = \frac{1}{(i-1)!(N-i)!} \sum_{P, P'} (-1)^P (-1)^{P'} A_{P(1),P'(1)} \dotsm A_{P(i-1),P'(i-1)} \phi_{P(i)} \phi_{P'(i)} B_{P(i+1),P'(i+1)} \dotsm B_{P(N),P'(N)} .
\end{equation}
Note that we did not explicitly write out the $z$ dependence of $\rho^{(i)}$, $A_{ij}$, $B_{ij}$, and $\phi_i$. Next, we introduce the permutations $P''$, defined by $P'=P'' \circ P$, and sum over $P$ and $P''$,
\begin{equation}
\rho^{(i)} = \frac{1}{(i-1)!(N-i)!} \sum_{P, P''} (-1)^{P''} \! A_{P(1),P'' \circ P(1)} \! \dotsm \! A_{P(i-1),P'' \circ P(i-1)} \phi_{P(i)} \phi_{P'' \circ P(i)} B_{P(i+1),P'' \circ P(i+1)} \! \dotsm \! B_{P(N),P'' \circ P(N)} .
\end{equation}
The order within the products of the $A$ and $B$ integrals is irrelevant. Hence, there are many equal terms in the above sum. In order to unite these terms, instead of summing over $P \in S_N$, we sum in the following over all decompositions $J+K+L=\underline N$ with $J=\{P(1), \dotsc, P(i-1)\}$, $K=\{P(i)\}$, $L=\{P(i+1), \dotsc, P(N)\}$, and $\underline N=\{1, \dotsc, N\}$. Then, since the order within the sets $J$ and $L$ is irrelevant, we have to multiply each term by $(i-1)!(N-i)!$ and obtain
\begin{equation}
\rho^{(i)} = \sum_{P''} \sum_{J+K+L=\underline N} (-1)^{P''} \prod_{j \in J} A_{j,P''(j)} \prod_{k \in K} \phi_k \phi_{P''(k)} \prod_{l \in L} B_{l,P''(l)} .
\end{equation}
Now, we use $B_{ij} = \delta_{ij} - A_{ij}$ to replace $B_{ij}$. One finds
\begin{equation}
\prod_{l \in L} B_{l,P''(l)} = \sum_{M+Q=L} (-1)^{|L|+|M|} \prod_{q \in Q} A_{q,P''(q)} .
\end{equation}
Here, we sum over all decompositions $M+Q=L$, where all elements of $M$ are mapped onto themselves by $P''$. Suppose $P''$ maps the elements $1$ and $2$ of $L$ onto themselves. Then, we can build the sets $M=\emptyset$, $\{1\}$, $\{2\}$, and $\{1,2\}$. Using this, we obtain, in the next step,
\begin{equation}
\rho^{(i)} = \sum_{P''} \sum_{J+K+M+Q=\underline N} (-1)^{P''+|L|+|M|} \prod_{j \in J} A_{j,P''(j)} \prod_{k \in K} \phi_k \phi_{P''(k)} \prod_{q \in Q} A_{q,P''(q)} .
\end{equation}
Next, we join the sets $J$ and $Q$ to form the set $R$, $J+Q=R$, and sum over all decompositions $K+M+R=\underline N$. There are $\binom{|R|}{|J|}$ different decompositions of $R$ into $J$ and $Q$. Moreover, we use
\begin{equation}
\sum_{P'' \in S_N} \sum_{K+M+R=\underline N} \dotsm = \sum_{K+M+R=\underline N} \sum_{P''' \in S_{K+R}} \dotsm
\end{equation}
to obtain
\begin{equation}
\rho^{(i)} = \sum_{K+M+R=\underline N} (-1)^{|L|+|M|} \binom{|R|}{|J|} \sum_{P''' \in S_{K+R}} (-1)^{P'''} \prod_{k \in K} \phi_k \phi_{P'''(k)} \prod_{r \in R} A_{r,P'''(r)} .
\end{equation}
In the next step, we use that any permutation $P''' \in S_{K+R}$ is either a composition of the form $P''' = \mathrm{id}_k \circ P''''$ or $P''' = P_{k,r} \circ P''''$, where $P'''' \in S_R$, $\mathrm{id}_k$ maps the element $k \in K$ on itself, and $P_{k,r}$ permutes the element $k \in K$ with one element $r \in R$. Therefore, we obtain
\begin{equation}
\rho^{(i)} \! = \! \sum_{K+M+R=\underline N} (-1)^{|L|+|M|} \binom{|R|}{|J|} \!\! \prod_{k \in K} \! \phi_k \Biggl\{ \! \phi_k \! \sum_{P'''' \in S_R} \! (-1)^{P''''} \! \prod_{r \in R} \! A_{r,P''''(r)} - \sum_{r \in R} \phi_r \! \sum_{P'''' \in S_R} \! (-1)^{P''''} \! \prod_{s \in R} \! A_{s,P_{k,r} \circ P''''(s)} \! \Biggr\} .
\end{equation}
Now, we want to sum over $j \in K$, $k = r \in R$, and $l=|R|$. Therefore, we express $|L|$, $|M|$, and $|J|$ by $i$, $j$, $k$, $l$, and $N$. It follows from the definitions of $J$ and $L$ that $|J|=i-1$ and $|L|=N-i$. Moreover, $|M|=N-1-l$, since $M = \underline N \setminus (K+R)$, $|K|=1$, and $|R|=l$. Which values can $j$, $k$, and $l$ assume? We can form the sets $K=\{1\}$, $\{2\}$, $\dotsc$, $\{N\}$, therefore $j=1,\dotsc,N$. $K$ and $R$ are disjoint, $K \cap R = \emptyset$, therefore $k = 1,\dotsc,N$ but $k \neq j$. Finally, $l=i-1$, $\dotsc$, $N-1$ since $l=|R|=|J|+|Q|=i-1+|Q|$ and $0 \leq |Q| \leq |L|=N-i$. We therefore obtain
\begin{equation}
\rho^{(i)} = (-1)^{i-1} \sum_{j=1}^N \sum_{l=i-1}^{N-1} (-1)^l \binom{l}{i-1} \phi_j \Biggl\{ \phi_j \sum_{p \in P(j,j,l)} \det A_p - \sum_{k \neq j = 1}^N \phi_k \sum_{p \in P(j,k,l)} \det A_p \Biggr\} .
\end{equation}
\end{widetext}
Here, we defined $P(j,j,l) = (\underline N \setminus \{j\})_l$, where $(M)_l$ is the set of all $l$-dimensional ordered subsets of $M$, and $P(j,k,l) = \left\{ m + \{k\} \; \text{with} \; m \in (\underline N \setminus \{j,k\})_{l-1} \right\}$, i.e., $P(j,k,l)$ is the set of all $l$-dimensional ordered subsets of $\underline N \setminus \{j\}$ that contain $k$. The matrix $A_p$ is defined by $A_p = (A_{ij})_{i \in p, j \in p}$ for $p \in P(j,j,l)$ and $A_p = (A_{ij})_{i \in p, j \in \overline p}$ for $p \in P(j,k,l)$, where $\overline p$ is obtained from $p$ by replacing $k$ by $j$. Additionally, we define $\sum_{p \in P(j,k,0)} \det A_p = \delta_{jk}$. Using the symmetry $A_{ij}=A_{ji}$, we finally obtain Eq.~(\ref{eq-rhoiz-3}).

\section{Exchange coefficients}
\label{app-Ji}

Here, we derive Eq.~(\ref{eq-Ji-2}) from Eq.~(\ref{eq-Ji}). We need for this purpose another formula for the single-particle densities,
\begin{eqnarray} \label{eq-rhoiz-5}
\rho^{(i)}(z) & = & \sum_{k = 0}^{N - i} \frac{(-1)^{N - i}}{k!} \, \binom{N-k-1}{i-1} \nonumber \\
& & \times \frac{d}{dz} \left[ \frac{\partial^k}{\partial\lambda^k} \det \Bigl( A(z) - \lambda \openone_N \Bigr) \right]_{\lambda = 0} ,
\end{eqnarray}
which is given in Ref.~\cite{Deuretzbacher08} and derived in Ref.~\cite{Deuretzbacher09}. Here, $A(z) = [A_{ij}(z)]_{i,j=1,\dotsc,N}$ and $\openone_N$ is the $N \times N$ identity matrix. By evaluating the integral~(\ref{eq-Ji}), we obtain
\begin{eqnarray}
J_i & = & \frac{N! \hbar^4}{m^2 g} \int_{z_1 < \dotsb < z_{i-1} < z_{i+1} < z_{i+2} < \dotsb < z_N} dz_1 \dotsi dz_{i-1} \nonumber \\
& & \times dz_{i+1} dz_{i+2} \dotsi dz_N \left| \frac{\partial \psi_F}{\partial z_i} \right|_{z_i=z_{i+1}}^2 .
\end{eqnarray}
The integrand is symmetric under permutations of $z_1, \dotsc, z_{i-1}$ and $z_{i+2}, \dotsc, z_N$. Therefore, we can extend the domain of integration to the domain $z_1, \dotsc, z_{i-1} < z_{i+1} < z_{i+2}, \dotsc, z_N$, divide by the factor $(i-1)!(N-i-1)!$, and get
\begin{eqnarray}
\mspace{-20mu} J_i & = & \frac{\hbar^4 N!}{m^2 g (i-1)! (N-i-1)!} \int_{-\infty}^{+\infty} dz_{i+1} \nonumber \\
& & \times \int_{-\infty}^{z_{i+1}} dz_1 \dotsi \int_{-\infty}^{z_{i+1}} dz_{i-1} \nonumber \\
& & \times \int_{z_{i+1}}^{+\infty} dz_{i+2} \dotsi \int_{z_{i+1}}^{+\infty} dz_N \left| \frac{\partial \psi_F}{\partial z_i} \right|_{z_i=z_{i+1}}^2 .
\end{eqnarray}
\begin{widetext}
Using the Laplace expansion along the $i$th row of the Slater determinant $\psi_F$ and the Leibniz formula for the minors,
\begin{equation}
\psi_F = \frac{1}{\sqrt{N!}} \sum_{j=1}^N (-1)^{i+j} \phi_j(z_i) \sum_{P \in S_{\underline N \setminus \{j\}}} (-1)^{P} \prod_{k=1}^{i-1} \phi_{P(k)}(z_k) \prod_{k=i}^{N-1} \phi_{P(k)}(z_{k+1}) ,
\end{equation}
we obtain (after renaming $z_N \rightarrow z_{N-1} \rightarrow \dotso \rightarrow z_{i+1} \rightarrow z_i$)
\begin{eqnarray}
J_i & = & \frac{\hbar^4}{m^2 g} \sum_{j=1}^N \sum_{k=1}^N (-1)^{j+k} \int_{-\infty}^{+\infty} dz_i \, \phi_j'(z_i) \phi_k'(z_i) \Biggl\{ \frac{1}{(i-1)!(N-1-i)!} \int_{-\infty}^{z_i} dz_1 \dotsi \int_{-\infty}^{z_i} dz_{i-1} \nonumber \\
& & \times \int_{z_i}^{+\infty} dz_{i+1} \dotsi \int_{z_i}^{+\infty} dz_{N-1} \sum_{P \in S_{\underline N \setminus \{j\}}} \sum_{P' \in S_{\underline N \setminus \{k\}}} (-1)^P (-1)^{P'} \prod_{l=1}^{N-1} \phi_{P(l)}(z_l) \phi_{P'(l)}(z_l) \Biggr\} .
\end{eqnarray}
The term in the braces resembles the $i$th single-particle density $\rho^{(i)}(z_i)$ of an $(N-1)$-particle system; see Eq.~(\ref{eq-rhoiz-4}). We therefore obtain, using Eq.~(\ref{eq-rhoiz-5}),
\begin{equation}
J_i = \frac{\hbar^4}{m^2 g} \sum_{j=1}^N \sum_{k=1}^N (-1)^{j+k} \int_{-\infty}^{+\infty} dz_i \, \phi_j'(z_i) \phi_k'(z_i) \sum_{l = 0}^{N-1-i} \frac{(-1)^{N-1-i}}{l!} \, \binom{N-l-2}{i-1} \frac{d}{dz_i} \left[ \frac{\partial^l}{\partial\lambda^l} \det \, \Bigl\langle A(z_i) - \lambda \openone_N \Bigr\rangle_{jk} \right]_{\lambda = 0} .
\end{equation}
Here, $A(z) = [A_{ij}(z)]_{i,j=1,\dotsc,N}$, $\openone_N$ is the $N \times N$ identity matrix, and $\langle \cdot \rangle_{jk}$ denotes the matrix operation of deleting the $j$th row and the $k$th column. After renaming $z_i \rightarrow z$, integrating by parts, and using the symmetry under the exchange $j \leftrightarrow k$, we obtain Eq.~(\ref{eq-Ji-2}).
\end{widetext}

\end{appendix}

\bibliographystyle{prsty}

\begin{thebibliography}{}
\bibitem{AncillaryFiles} See the ancillary files for the {\footnotesize MATHEMATICA} notebook used to calculate the results of this paper, a detailed derivation of the formula for the one-body density matrix elements, Eq.~(\ref{eq-rhoijzz'-2}), and a table with the exchange coefficients of up to 60 harmonically trapped particles. A less efficient method for calculating the exchange coefficients that works for up to 30 particles was given in the 2nd version of this manuscript, \arxiv{1602.06816v2}.
\end{thebibliography}

\end{document}